\pdfoutput=1
\RequirePackage{ifpdf}
\ifpdf 
\documentclass[pdftex]{sigma}
\else
\documentclass{sigma}
\fi

\numberwithin{equation}{section}

\newtheorem{Theorem}{Theorem}[section]
\newtheorem*{Theorem*}{Theorem}
\newtheorem{Corollary}[Theorem]{Corollary}
\newtheorem{Lemma}[Theorem]{Lemma}
\newtheorem{Proposition}[Theorem]{Proposition}

\theoremstyle{definition}
\newtheorem{Definition}[Theorem]{Definition}

\newtheorem{Example}[Theorem]{Example}
\newtheorem{Remark}[Theorem]{Remark}

\usepackage{mathtools}
\usepackage{nicematrix}
\usepackage{bm}
\usepackage{tikz}
\graphicspath{{figures/} }
\usepackage{adjustbox}

\usepackage[labelfont=bf,labelsep=period,font=small]{caption}
\usepackage[subrefformat=parens]{subcaption}

\begin{document}

\allowdisplaybreaks

\newcommand{\arXivNumber}{2510.12905}

\renewcommand{\PaperNumber}{082}

\FirstPageHeading

\ShortArticleName{Constructing Solutions of Simplex Equations from Polygon Equations}

\ArticleName{Constructing Solutions of Simplex Equations\\ from Polygon Equations}

\Author{Serban Matei MIHALACHE~$^{\rm a}$ and Tomoro MOCHIDA~$^{\rm b}$}

\AuthorNameForHeading{S.M.~Mihalache and T.~Mochida}

\Address{$^{\rm a)}$~Graduate School of Mathematical Sciences, The University of Tokyo,\\
\hphantom{$^{\rm a)}$}~3-8-1 Komaba, Meguro, Tokyo, 153-8914, Japan}
\EmailD{\mail{mateimihalache@g.ecc.u-tokyo.ac.jp}}

\Address{$^{\rm b)}$~Mathematical Institute, Tohoku University,\\
\hphantom{$^{\rm b)}$}~6-3, Aoba, Aramaki-aza, Aoba-ku, Sendai, 980-8578, Japan}
\EmailD{\mail{tomorou.mochida.r5@dc.tohoku.ac.jp}}

\ArticleDates{Received January 27, 2026, in final form August 12, 2026; Published online August 25, 2026}

\Abstract{We study polygon equations and their connections to simplex equations, which generalize the pentagon and Yang--Baxter equations, respectively. First, we show that certain ``commutative'' pairs of solutions of (dual) polygon equations give rise to solutions of higher-order polygon equations. Next, we define an explicit compatibility condition between solutions of the $n$-gon and dual $n$-gon equations and use it to construct solutions of the $(n-2)$- and $(n-1)$-simplex equations. This extends earlier work by Kashaev--Sergeev and Dimakis--M\"uller-Hoissen.}


\Classification{15A69; 16T25; 16T05; 57Q15}

\section{Introduction}

The Yang--Baxter equation~\cite{baxter1972partition, yang1967some},
\begin{align*}
 R_{12}R_{13}R_{23}=R_{23}R_{13}R_{12},
\end{align*}
originally introduced in the study of integrable models in statistical mechanics, is now fundamental to mathematical physics, topology, and many other areas of mathematics. A.B.~Zamolodchikov~\cite{zamolodchikov1980tetrahedra,zamolodchikov1981tetrahedron} considered a three-dimensional analog of the Yang--Baxter equation, known as the tetrahedron equation.
Later, higher-dimensional analogs of these equations, called the simplex equations, were defined in \cite{bazhanov1982conditions}, including the Yang--Baxter equation as the 2-simplex equation, Zamolodchikov's tetrahedron equation as the 3-simplex equation, and so on. See, for example, \cite{bardakov2024set, carter1996formulations,dimakis2015simplex, jimbo1989introduction,
korepanov2016cohomologies,maillet1989tetrahedron} for more details.

Another important equation with a flavor similar to the Yang--Baxter equation is the pentagon (5-gon) equation,
\begin{align*}
 T_{12}T_{13}T_{23}=T_{23}T_{12},
\end{align*}
which again appears as a fundamental equation in mathematical physics and topology. Subsequently, Dimakis and M\"uller-Hoissen~\cite{dimakis2015simplex} defined the class of polygon equations. Just as simplex equations generalize the Yang--Baxter and tetrahedron equations, polygon equations generalize the pentagon equation. In \cite{muller2024structure}, M\"uller-Hoissen explored (set-theoretic) polygon equations, investigating the relationship between neighboring polygon equations and how a solution of the $k$-gon equation lifts or descends to a solution of the $(k\pm1)$-gon equation. Also, in topology, polygon equations play a key role in constructing state sum invariants of manifolds from their triangulations~\cite{kashaev2018simple,korepanov2024odd, korepanov2017hexagon}. In particular, they guarantee invariance under a certain Pachner move.

As for the relationship between simplex equations and polygon equations, there have been several attempts to construct solutions of simplex equations from solutions of polygon equations. In \cite{kashaev1998pentagon}, R.M. Kashaev and S.M. Sergeev showed that whenever a solution of the pentagon equation and a solution of its dual satisfy a certain compatibility condition, called the ``ten-term relation'', they give rise to solutions of the tetrahedron (3-simplex) equation and the 4-simplex equation. See also \cite{kassotakis2023matrix}. Later in \cite{dimakis2015simplex}, Dimakis and M\"uller-Hoissen associated simplex equations with higher Bruhat orders and polygon equations with higher Tamari orders. Then, by considering the ``three color decomposition'', which is a decomposition of the higher Bruhat order into the higher Tamari order, its dual, and what they called the mixed order, they constructed a solution of the $(n-1)$-simplex equation from a compatible pair of solutions of the $n$-gon equation and its dual. This generalizes Kashaev and Sergeev's construction of a solution of the 4-simplex equation from solutions of the pentagon equation and its dual. However, as pointed out in \cite[Section~6]{dimakis2015simplex}, they were unable to fully explore the relationship between solutions of the 3-simplex equation and those of the pentagon equation. Also, the compatibility condition between solutions of the $n$-gon equation and its dual was not made explicit, by which we mean that although one can write the compatibility condition for each $n$ individually, it is still unclear how to express it uniformly. A similar study, particularly for the class of odd-gon equations, also appears in Dimakis and~Korepanov's work \cite{dimakis2021grassmannian}.

In this paper, we study polygon and simplex equations. We begin by exploring polygon equations within a general framework. We show that if two solutions of (dual) polygon equations are ``commutative'' in the sense of Section~\ref{sec:stacking_solutions}, their (partial) composition and tensor product yield a~solution of a higher-order polygon equation. In the second part, we explore the relationship between simplex equations and polygon equations. We explicitly define the compatibility condition above for solutions of the $n$-gon and dual $n$-gon equations and observe how these two solutions give rise to a solution of the $(n-2)$-simplex equation and, for completeness, to a~solution of the $(n-1)$-simplex equation. We provide direct proofs of these connections for both even- and odd-gon cases.

There is also an observation of the relationship between the pentagon equation and the Yang--Baxter (2-simplex) equation in \cite{evripidou2024quadrirational, kashaev1995heisenberg,suzuki2018universal}. It is natural to ask for a generalization of this result, namely the relationship between the $n$-gon equation and the $(n-3)$-simplex equation, and we leave this for future work.

The rest of the paper is organized as follows: In Section~\ref{sec:representing_linear_maps}, we develop graphical techniques for representing linear maps. We employ them throughout the paper to visualize simplex and polygon equations and to prove most of the theorems and propositions. In Section~\ref{sec:polygon_equations}, we define polygon and dual polygon equations using both multi-index notation and graphical representations. We then prove some properties of solutions of (dual) polygon equations. In Section~\ref{sec:stacking_solutions}, we consider constructing a new solution by (partially) composing two solutions of lower polygon equations that are commutative in a certain sense. As a corollary, we prove all the conjectures posed in \cite{muller2024structure}. Examples of solutions of polygon equations are given in Section~\ref{sec:examples_polygon_eq}. In Section~\ref{sec:simplex_from_polygon}, we define the mixed relation between solutions of the $n$-gon and the dual $n$-gon equations. Out of such a mixed pair of solutions, we construct solutions of the $(n-1)$- and $(n-2)$-simplex equations. This generalizes the construction of solutions for both 3-gon and 4-gon equations given by Kashaev and Sergeev~\cite{kashaev1998pentagon}. In Appendix~\ref{sec:pachnerpolygon}, we explain the relationship between Pachner moves in PL topology and polygon equations. Diagrammatic constructions of (non-constant) polygon and simplex equations are presented. Appendix~\ref{sec:nonconstantthm} considers a generalization of the mixed relation and Theorems~\ref{thm:fromnton-1} and \ref{thm:fromnton-2} in the non-constant setting. Appendix~\ref{sec:equivalence} compares the definition of polygon equations given in \cite{dimakis2015simplex} with ours.

Throughout the paper, we fix a field $\Bbbk$, and vector spaces, linear maps, tensor products, duals, etc., are considered to be over $\Bbbk$ unless otherwise stated. Note that many of the arguments in this paper can be generalized to polygon and simplex equations in a symmetric monoidal category.

\section{Representing linear maps}\label{sec:representing_linear_maps}

In this section, we describe how a linear map indexed by a multi-index acts on a tensor product of vector spaces and introduce graphical representations of such linear maps. This setup will be used throughout the paper to define polygon and simplex equations and to prove most of our results.

Let $V$ be a vector space, $F\colon V_1\otimes \cdots \otimes V_k\to V_1\otimes \cdots \otimes V_l$ be a linear map, where each $V_i$ is just a copy of $V$, and $n\geq k$ be an integer.
Let $(a,b)$ be a pair of multi-indices (i.e., row vectors)
\begin{align*}
 a=[ a_{1},a_{2},\dots, a_{k} ]\qquad \text{and} \qquad b=[ b_{1},b_{2},\dots, b_{l} ]
\end{align*}
such that $1\le a_1<a_2<\cdots<a_k\le n$ and $1\le b_1<b_2<\cdots< b_l\le n-k+l$.

Then $F_{(a,b)}\colon V_1\otimes \cdots \otimes V_n\to V_1\otimes \cdots \otimes V_{n-k+l}$ is the map obtained by letting $F$ act non-trivially only on the positions in $V^{\otimes n}$ specified by $a$ and placing its outputs on the positions in $V^{\otimes (n-k+l)}$ specified by $b$.
Formally, \begin{align*}
 F_{(a,b)}\coloneqq\tau_{1,b_1}^{-1} \tau_{2,b_2}^{-1} \cdots \tau_{l,b_l}^{-1}\bigl(F\otimes\operatorname{id}_V^{\otimes{(n-k)}}\bigr) \tau_{k,a_k} \cdots \tau_{2,a_2} \tau_{1,a_1},
\end{align*}
where $\tau_{i,j}$ is a map that sweeps the $j$-th factor to the $i$-th position if $i\ne j$, i.e.,
\begin{align*}
 \tau_{i,j}(\cdots \otimes x_i \otimes \cdots \otimes x_{j-1} \otimes x_j \otimes x_{j+1}\otimes \cdots ) \coloneqq \cdots \otimes x_{j} \otimes x_{i} \otimes \cdots \otimes x_{j-1} \otimes x_{j+1}\otimes\cdots,
\end{align*}
and is the identity if $i=j$.

We mainly use the following notation:
\begin{itemize}\itemsep=0pt
 \item For a linear map $F\colon V^{\otimes k}\to V^{\otimes k}$ ($l=k$ case) and a multi-index $a$, $F_{a}$ denotes the map $F_{(a,a)}$. This is just the standard notation specifying which factors of the tensor product the map $F$ acts on.
 \item For a linear map $F\colon V^{\otimes k}\to V^{\otimes k+1}$ ($l=k+1$ case), if multi-indices $a$ and $b$ are of the forms
 \begin{equation*}
 a=[ a_{1},a_{2},\dots, a_{k} ] \qquad\text{and}\qquad b=[ a_{1},a_{2},\dots, a_{k}, a_{k}+1 ],
 \end{equation*}
 respectively, then we write $F_a$ for $F_{(a,b)}$.
 \item For a linear map $F\colon V^{\otimes k}\to V^{\otimes k-1}$ ($l=k-1$ case), if multi-indices $a$ and $b$ are of the forms
 \begin{equation*}
 a=[ a_{1},a_{2},\dots, a_{k-1},a_{k-1}+1 ]\qquad\text{and}\qquad b=[ a_{1},a_{2},\dots, a_{k-1} ],
 \end{equation*}
 respectively, then we write $F_b$ for $F_{(a,b)}$.
\end{itemize}

Next, we introduce graphical representations of linear maps.
Since our definition and treatment of polygon equations and simplex equations rely heavily on these graphical representations, we give a detailed explanation of them here.

Given a linear map $T\colon V^{\otimes k}\to V^{\otimes k}$, we will represent $T$ by an oriented (blue) line with $k$ dots:
\begin{equation*}
 \adjustimage{valign=m,scale=0.8}{ex_of_linear_map_n.pdf}
\end{equation*}
The $i$-th vertical line represents the $i$-th input/output of the linear map $T$, where the incoming vertical line represents the input and the outgoing vertical line represents the output.

Similarly, given linear maps $Q\colon V^{\otimes k-1}\to V^{\otimes k}$ and $U\colon V^{\otimes k}\to V^{\otimes k-1}$, we will draw~$Q$ and~$U$ as
\begin{equation*}
 \adjustimage{valign=m,scale=0.8}{ex_of_linear_map_2_n.pdf}
\end{equation*}

The (partial) composition of linear maps is represented by stacking the diagrams and connecting the output legs to the input legs.
For example, the linear map $(\operatorname{id}_V \otimes U)\circ(Q \otimes \operatorname{id}_V )$ acting on $V^{\otimes k}$ is represented as
\begin{equation*}
 \adjustimage{valign=m, scale=0.8}{ex_of_linear_map_3_n.pdf}
\end{equation*}

In the next section, we will define the main object of this paper, the $n$-gon equations.
These equations will be given by specifying indexing, and then translated into the graphical representation introduced in this section.
To help the reader understand the graphical representation of $n$-gon equations given in Section \ref{sec:polygon_equations}, we give examples of the $5$-gon (pentagon) and $6$-gon (hexagon) equations.

For a map $T\colon V^{\otimes 2}\to V^{\otimes 2}$, the $5$-gon (pentagon) equation is given by
\begin{equation*}
T_{12}T_{13}T_{23}=T_{23}T_{12}
\end{equation*}
in $\operatorname{End}\bigl(V^{\otimes 3}\bigr)$, and the map $T$ is called a solution of the 5-gon equation.
The graphical representation of this equation is given as follows:
\begin{equation*}
 \adjustimage{valign=m, scale=0.8}{5-gon_flat.pdf}
\end{equation*}
These diagrams are read from bottom to top, unless stated otherwise.
The map $T_{13}$ appearing in the pentagon equation acts on the vector spaces in the first and third positions, which are not consecutive.
So, to draw the diagram, one has to ``stretch'' the map $T$.
In order to avoid this ``stretching'', we regard the diagram as embedded in $\mathbb{R}^3$, which is depicted on the left-hand side of the figure below:
\begin{equation*}
 \adjustimage{valign=m, scale=0.9}{5-gon_projection.pdf}
\end{equation*}
If we project it onto the $yz$-plane (with a small perturbation), we recover the original diagram of the 5-gon equation.
We will instead project this embedded diagram in $\mathbb{R}^3$ onto the $xy$-plane. The result is depicted on the right-hand side of the above figure.
In order to make clear both (1) which outputs are used as inputs of each map and (2) the order in which the maps are composed, we depict this projection onto the $xy$-plane at a slight angle. This allows us to recover the original diagram from the projected one.
As we will see in Section \ref{sec:polygon_equations}, the resulting diagram is easily generalized to the general form of the $n$-gon equation.

Another example is the $6$-gon (hexagon) equation.
For a map $T\colon V^{\otimes 2}\to V^{\otimes 3}$, the $6$-gon (hexagon) equation is given by
\begin{equation*}
T_{12}T_{13}T_{23}=T_{35}T_{24}T_{12},
\end{equation*}
in $\operatorname{Hom}\bigl(V^{\otimes 3}, V^{\otimes 6}\bigr)$ and the map $T$ is called a solution of the 6-gon equation.
The graphical representation of this diagram is given as follows:
\begin{equation*}
 \adjustimage{valign=m, scale=0.8}{example_for_hexagon_1.pdf}
\end{equation*}
We can apply the same procedure as in the 5-gon case above. Namely, we embed the diagram into $\mathbb{R}^3$ and project it onto the $xy$-plane to get a nice graphical representation:
\begin{equation*}
 \adjustimage{valign=m, scale=0.78}{6-gon_projection.pdf}\adjustbox{raise=-2ex}{}
\end{equation*}

We finally introduce the left and right partial traces. Suppose that $V$ is finite-dimensional. For $F\in\operatorname{Hom}\bigl(V^{\otimes i},V^{\otimes j}\bigr)$, the left (resp.\ right) partial trace $\operatorname{tr}_l(F) \text{ (resp.\ $\operatorname{tr}_r(F)$)}\in\operatorname{Hom}\bigl(V^{\otimes {i-1}},\allowbreak V^{\otimes {j-1}}\bigr)$ is defined by taking the trace over the leftmost (resp. rightmost) input and output, or graphically, (just as in the case of string diagrams) by a circle looping at the leftmost (resp. rightmost) dot:
\begin{equation*}
 \includegraphics[scale=0.8]{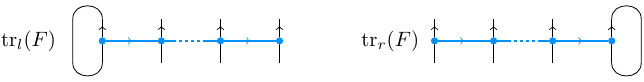}
\end{equation*}

\section{Polygon equations}\label{sec:polygon_equations}

In this section, we define the polygon and dual polygon equations. The polygon and dual polygon equations are a family of equations indexed by an integer $n\geq 3$, where the equations associated to a specific $n$ are called the $n$-gon and dual $n$-gon equations.

Originally, the definitions were given in \cite{dimakis2015simplex} in terms of the higher Tamari order.
Instead, we define them in tensor notation, namely by directly specifying the factors on which each operator in the equation acts.
We also give definitions in graphical representation. See Appendix~\ref{sec:pachnerpolygon} for a geometric background of these equations.

\begin{Definition}[odd-gon equation]\label{def:oddgon}
 For an integer $k\geq 1$, the $(2k+1)$-gon equation for a~linear~map $T\colon V^{\otimes k}\to V^{\otimes k}$ is an equation in \smash{$\operatorname{End}\bigl(V^{\otimes k(k+1)/2}\bigr)$} of the form
 \begin{equation*}
 T_{a_1}T_{a_2}\cdots T_{a_{k+1}} = T_{b_{k}}T_{b_{k-1}}\cdots T_{b_1},
 \end{equation*}
 where $a_i=[a_{i,1},a_{i,2},\dots, a_{i,k}]$ and $b_j=[b_{j,1},b_{j,2},\dots, b_{j,k}]$ are multi-indices indicating which fac\-tors of \smash{$V^{\otimes k(k+1)/2}$} the operator $T$ acts on and are given by the following matrices~$A_{2k+1}$ and~$B_{2k+1}$, respectively:
 \begin{align*}
 & A_{2k+1} = \begin{bmatrix}
 a_1\\
 a_2\\
 \vdots\\
 a_{k+1}
 \end{bmatrix}
 = \begin{bNiceArray}{c|w{c}{1cm}w{c}{0.5cm}w{c}{1cm}}[margin]
 1 & 2 & \cdots & k \\\hline
 1 & \Block{4-3}{A_{2k-1}+[\,k\,]_{k,k-1}} \\
 2\\
 \vdots \\
 k \\
 \end{bNiceArray},\\
 & B_{2k+1} = \begin{bmatrix}
 b_1\\
 b_2\\
 \vdots\\
 b_{k}
 \end{bmatrix}
 = \begin{bNiceArray}{c|w{c}{1cm}w{c}{0.5cm}w{c}{1cm}}[margin]
 1 & 2 & \cdots & k \\\hline
 2 & \Block{3-3}{B_{2k-1}+[\,k\,]_{k-1,k-1}} \\
 \vdots \\
 k \\
 \end{bNiceArray},
 \end{align*}
 with $A_3 = [ 1,1 ]^{\mathsf{T}}$ and $B_3 = [ 1 ]$, where $[ r ]_{i,j}$ denotes the $i\times j$ matrix whose entries are all $r$.

 Alternatively, the $(2k+1)$-gon equation can be defined graphically by
 \begin{equation*}
 \adjustimage{valign=b}{odd-gon.pdf}
 \end{equation*}
 (it is read from bottom to top).
\end{Definition}

We can also define the dual version of Definition~\ref{def:oddgon}.

\begin{Definition}[dual odd-gon equation]\label{def:dual oddgon}
 For an integer $k\geq 1$, the dual $(2k+1)$-gon equation for a linear map $T\colon V^{\otimes k}\to V^{\otimes k}$ is an equation in \smash{$\operatorname{End}\bigl(V^{\otimes k(k+1)/2}\bigr)$} of the form:
 \begin{equation*}
 T_{a_{k+1}}T_{a_k}\cdots T_{a_1} = T_{b_1}T_{b_2}\cdots T_{b_k},
 \end{equation*}
 where the multi-indices $a_i$ and $b_j$ are the same as in Definition~\ref{def:oddgon}.

 Since the dual version is obtained by just reversing the order of compositions, the graphical representation of the dual $(2k+1)$-gon equation is given by reversing the composition arrows in the diagram of the $(2k+1)$-gon equation, or, in other words, reading the diagram from top to bottom.
\end{Definition}

\begin{Definition}[even-gon equation]\label{def:evengon}
 For an integer $k\geq 2$, the $2k$-gon equation for a linear map $T\colon V^{\otimes k-1}\to V^{\otimes k}$ is an equation in \smash{$\operatorname{Hom}\bigl(V^{\otimes k(k-1)/2}, V^{\otimes k(k+1)/2}\bigr)$} of the form:
 \begin{equation*}
 T_{a_1}T_{a_2}\cdots T_{a_{k}} = T_{b_{k}}T_{b_{k-1}}\cdots T_{b_1},
 \end{equation*}
 where the multi-indices $a_i=[a_{i,1},a_{i,2},\dots, a_{i,k-1}]$ and $b_j=[b_{j,1},b_{j,2},\dots, b_{j,k-1}]$ are given by the following matrices $A_{2k}$ and $B_{2k}$, respectively:
 \begin{equation*}
 A_{2k}=\begin{bmatrix}
 a_1\\
 a_2\\
 \vdots\\
 a_{k}
 \end{bmatrix}:=A_{2k-1},\quad
 B_{2k} = \begin{bmatrix}
 b_1\\
 b_2\\
 \vdots\\
 b_{k}
 \end{bmatrix}
 := \text{$B_{2k+1}$ with the last column deleted.}
 \end{equation*}

 Alternatively, the $2k$-gon equation can be defined graphically by
 \begin{equation*}
 \adjustimage{valign=b}{even-gon.pdf}
 \end{equation*}
 (it is read from bottom to top).
\end{Definition}

Similarly, we can define the dual version.
\begin{Definition}[dual even-gon equation]
 For an integer $k\geq 2$, the dual $2k$-gon equation for a~linear map $T\colon V^{\otimes k}\to V^{\otimes k-1}$ is an equation in \smash{$\operatorname{Hom}\bigl(V^{\otimes k(k+1)/2}, V^{\otimes k(k-1)/2}\bigr)$} of the form
 \begin{equation*}
 T_{a_k}T_{a_{k-1}}\cdots T_{a_1} = T_{b_1}T_{b_2}\cdots T_{b_k},
 \end{equation*}
 where the multi-indices $a_i$ and $b_j$ are the same as in Definition~\ref{def:evengon}.

 Again, since the dual version is obtained by just reversing the order of compositions, the diagram of the dual $2k$-gon equation is obtained by reversing the composition arrows in the diagram of the $2k$-gon equations, or by reading the diagram from top to bottom.
\end{Definition}

We have defined each type of polygon equation in two ways: tensor notation and graphical notation. Comparing these, the tensor notation arranges the input vectors of the (dual) polygon equation linearly and specifies the action of each $T$ by a multi-index, whereas the graphical notation places these input vectors on the $\mathbb{R}^2$ plane and connects them according to the action of $T$. This not only makes the structure of the equation much easier to understand but also, as will be seen later, allows us to treat the odd-gon and even-gon cases in a unified way. For this reason, graphical methods will be heavily used in arguments involving polygon (and simplex) equations throughout the paper.

\begin{Example}
 The $n$-gon equations for small $n$ are given as follows:
\begin{align*}
 &\begin{aligned}
 &\text{{3-gon:}}\ T_{1}T_{1}=T_{1}, & \text{{4-gon:}}&\quad T_{1}T_{1}=T_{2}T_{1},\\
 &\text{{5-gon:}}\ T_{12}T_{13}T_{23}=T_{23}T_{12}, & \text{{6-gon:}}&\quad T_{12}T_{13}T_{23}=T_{35}T_{24}T_{12},\\
 &\text{{7-gon:}}\ T_{123}T_{145}T_{246}T_{356}=T_{356}T_{245}T_{123},\quad & \text{{8-gon:}}&\quad T_{123}T_{145}T_{246}T_{356}=T_{479}T_{368}T_{256}T_{123},
 \end{aligned}\\
 &\begin{aligned}
 &\text{{9-gon:}}\ T_{1234}T_{1567}T_{2589}T_{368,10}T_{479,10}=T_{479,10}T_{3689}T_{2567}T_{1234},\\
 &\text{{10-gon:}}\ T_{1234}T_{1567}T_{2589}T_{368,10}T_{479,10}=T_{59,12,14}T_{48,11,13}T_{37,10,11}T_{2678}T_{1234}.
 \end{aligned}
\end{align*}
\end{Example}

\begin{Remark}\label{rem:translation}\quad
\begin{enumerate}\itemsep=0pt
 \item Note that we only treat the so-called constant polygon equations, where the maps $T$ appearing in the equation are all the same and act on copies of the same vector space~$V$.
 In general, one can consider the non-constant version, in which maps $T$ appearing in the equation, as well as vector spaces $V$ on which they act, are allowed to be different. See~\cite{dimakis2015simplex} for a general treatment.
 \item This definition of polygon equations looks slightly different from the definition given in \cite[Section~4.2]{dimakis2015simplex}, but they are indeed equivalent. The details are given in Appendix~\ref{sec:equivalence}.
 \item The tensor notation for odd-gon equations is also given in \cite[Proposition~3]{dimakis2021grassmannian}. Since their notation satisfies the recurrence formula in Definition~\ref{def:oddgon}, it defines the same family of equations as ours.
\end{enumerate}
\end{Remark}

We list some useful properties of solutions of polygon equations, many of which are similar to the properties of solutions of simplex equations \cite{bardakov2024set}.
\begin{Proposition}\label{prop:usefulpolygon}\quad
 \begin{enumerate}\itemsep=0pt
 \item[$(1)$] A~map $T$ is a solution of the $(2k+1)$-gon equation if and only if $T^{-1}$ $($if it exists$)$ is a~solution of the dual $(2k+1)$-gon equation
 \item[$(2)$] For a solution of the $($dual$)$ $(2k+1)$-gon equation $T$ and any $\phi\in\operatorname{Aut}(V)$, the map $(\phi^{-1})^{\otimes k}\circ T\circ\phi^{\otimes k}$ is a solution of the $($dual$)$ $(2k+1)$-gon equation.

 For a solution of the $2k$-gon (resp. dual $2k$-gon) equation $T$ and any $\phi\in\operatorname{Aut}(V)$, the map $(\phi^{-1})^{\otimes k}\circ T\circ\phi^{\otimes (k-1)}$ $($resp.\ $\bigl(\phi^{-1}\bigr)^{\otimes (k-1)}\circ T\circ\phi^{\otimes k})$ is a solution of the $2k$-gon $($resp.\ dual $2k$-gon$)$ equation.
 \item[$(3)$] A map $T$ is a solution of the $(2k+1)$-gon equation if and only if $\bar{\sigma}_k T\bar{\sigma}_k $ is a solution of the dual $(2k+1)$-gon equation, where $\bar{\sigma}_k\colon V^{\otimes k}\to V^{\otimes k}$ is an inversion given by
 \begin{equation}\label{eq:inversion}
 \bar{\sigma}_k(x_1\otimes x_2\otimes\cdots\otimes x_k) \coloneqq x_k\otimes x_{k-1}\otimes \cdots\otimes x_1.
 \end{equation}
 \item[$(4)$] A~map $T$ is a solution of the $2k$-gon $($resp.\ dual $2k$-gon$)$ equation if and only if $\bar{\sigma}_{k}T\bar{\sigma}_{k-1}$ $($resp.\ $\bar{\sigma}_{k-1}T\bar{\sigma}_{k})$ is a solution of the $2k$-gon $($resp.\ dual $2k$-gon$)$ equation.

 \item[$(5)$] If $T$ is an invertible solution of the $($dual$)$ $(2k+1)$-gon equation over a finite-dimensional vector space $V$, then $\dim(V)^{-1}\operatorname{tr}_l(T)$ and $\dim(V)^{-1}\operatorname{tr}_r(T)$ are solutions of the $($dual$)$ $(2k-1)$-gon equation.
 \end{enumerate}
\end{Proposition}

\begin{proof}
 (1), (2) They are obvious from the definition of the (dual) $n$-gon equation.

 (3) For a multi-index $a_i=[a_{i,1},a_{i,2},\dots, a_{i,k}]$, set $\bar{a}_i\coloneqq [a_{i,k},a_{i,k-1},\dots, a_{i,1}]$.
 Then it is easy to see that $T_{\bar{a}_i}=(\bar{\sigma}T\bar{\sigma})_{a_{i}}$.
 If we substitute $\bar{\sigma}T\bar{\sigma}$ into the polygon equation, we get
 \begin{equation*}
 T_{\bar{a}_1}T_{\bar{a}_2}\cdots T_{\bar{a}_{k+1}} = T_{\bar{b}_{k}}T_{\bar{b}_{k-1}}\cdots T_{\bar{b}_1}.
 \end{equation*}
 The diagram representing this equation is obtained from that of the odd-gon equation in Definition~\ref{def:oddgon} by reversing the orientation of each line.
 If we rotate both left-hand and right-hand sides of the diagram by 90 degrees counter-clockwise, and then take mirror images of both left-hand and right-hand sides of the diagram across a vertical line, the result is the dual $(2k+1)$-equation.

 (4) It can be proved in a similar way to~(3).

 (5) We show this for the 9-gon equation ($k=4$). The general case can be proved in the exact same way. The proof proceeds using the following graphical calculus:
 \begin{equation*}
 \adjustimage{valign=m}{partial_trace.pdf}
 \end{equation*}
 (recall that the loop represents the partial trace). We compose with $T_{a_1}^{-1}$ $\bigl(=T_{b_1}^{-1}\bigr)$ (shown in red) and take the partial trace over the position $a_1$ $(=b_1)$ on both sides of the 9-gon equation, yielding the two middle diagrams. By the cyclicity of the trace, we then cancel $T_{a_1}$ and $T_{b_1}$ against $T^{-1}$ and obtain the equality between the leftmost and rightmost diagrams. Thus if we normalize $\operatorname{tr}_l(T)$ by $\dim(V)^{-1}$, the proposition follows. For the $\operatorname{tr}_r(T)$ by $\dim(V)^{-1}$ case, replace~$a_1$~$(=b_1)$ with $a_{k+1}$ $(=b_k)$.
\end{proof}

\section{Stacking solutions}\label{sec:stacking_solutions}

For a linear map $F\colon V^{\otimes k}\to V^{\otimes l}$, let $F^{[n]}\colon \bigl(V^{\otimes n}\bigr)^{\otimes k} \to \bigl(V^{\otimes n}\bigr)^{\otimes l}$ be the linear map defined by
\begin{equation*}
\bigl(V^{\otimes n}\bigr)^{\otimes k}\cong \bigl(V^{\otimes k}\bigr)^{\otimes n} \xrightarrow{F^{\otimes n}} \bigl(V^{\otimes l}\bigr)^{\otimes n}\cong \bigl(V^{\otimes n}\bigr)^{\otimes l}.
\end{equation*}
Explicitly, for a pure tensor $(v_{1,1}\otimes\cdots\otimes v_{1,n})\otimes\cdots\otimes
(v_{k,1}\otimes\cdots\otimes v_{k,n})$, we first rearrange it as $(v_{1,1}\otimes\cdots\otimes v_{k,1})\otimes\cdots\otimes
(v_{1,n}\otimes\cdots\otimes v_{k,n})$. We then apply $F^{\otimes n}$ and write the resulting tensor formally as $\sum
(w_{1,1}\otimes\cdots\otimes w_{l,1})\otimes\cdots\otimes
(w_{1,n}\otimes\cdots\otimes w_{l,n})$. We finally rearrange it back to $\sum
(w_{1,1}\otimes\cdots\otimes w_{1,n})\otimes\cdots\otimes
(w_{l,1}\otimes\cdots\otimes w_{l,n})$.

A linear map $G\colon V^{\otimes i}\to V^{\otimes j}$ is said to commute with $F$, written as $F \leftrightarrow G$, if
\begin{equation}
G^{\otimes l}\circ F^{[i]} = F^{[j]}\circ G^{\otimes k}.\label{eq:commute}
\end{equation}
Note that this condition is equivalent to $F^{\otimes j}\circ G^{[k]} = G^{[l]}\circ F^{\otimes i}$. Thus, the relation $F \leftrightarrow G$ is symmetric in $F$ and $G$.
When $F$ is a product on $V$, i.e., $F\colon V^{\otimes 2}\to V$, then $F\leftrightarrow G$ is precisely the condition that $G$ be an algebra map.

For example, in the case of $k=l$ and $i=j$, the diagram of \eqref{eq:commute} is given by
\begin{equation*}
 \adjustimage{valign=m}{commuting_relation.pdf}
\end{equation*}
where $F$ is represented in blue, and $G$ in red.

Let $T^{(n)}$ and $T^{\prime(n)}$ denote solutions of the $n$-gon equation, and let $S^{(n)}$ and $S^{\prime(n)}$ denote solutions of the dual $n$-gon equation.

\begin{Proposition}\label{prop:commpolygon}
 Let $n$ and $k$ be any positive integers.
 Then the following statements hold for the partial composition and tensor product:
 \begin{enumerate}\itemsep=0pt
 \item[$(1)$] If $T^{(2k+1)}\leftrightarrow T^{\prime(n)}$, then
 $
 \begin{cases}
 \text{$T^{(2k+1)}\circ_l T^{\prime(n)}$ is a solution of the $(n+2k-2)$-gon equation,}\\
 \text{$T^{(2k+1)}\otimes T^{\prime(n)}$ is a solution of the $(n+2k)$-gon equation.}
 \end{cases} $\!\!
 \item[$(2)$] If $S^{(2k+1)}\leftrightarrow S^{\prime(n)}$, then
 $
 \begin{cases}
 \text{$S^{(2k+1)}\circ_r S^{\prime(n)}$ is a solution of the dual}\\[-1mm]
 \qquad \text{$(n+2k-2)$-gon equation,}\\
 \text{$S^{(2k+1)}\otimes S^{\prime(n)}$ is a solution of the dual $(n+2k)$-gon equation.}
 \end{cases} $\!\!\!\!
 \item[$(3)$] If $S^{(2k)}\leftrightarrow T^{(n)}$, then
 $
 \begin{cases}
 \text{$S^{(2k)}\circ_lT^{(n)}$ is a solution of the dual $(n+2k-3)$-gon equation,}\\
 \text{$S^{(2k)}\otimes T^{(n)}$ is a solution of the dual $(n+2k-1)$-gon equation.}
 \end{cases}
 $
 \item[$(4)$] If $T^{(2k)}\leftrightarrow S^{(n)}$, then
 $
 \begin{cases}
 \text{$T^{(2k)}\circ_rS^{(n)}$ is a solution of the $(n+2k-3)$-gon equation,}\\
 \text{$T^{(2k)}
 \otimes S^{(n)}$ is a solution of the $(n+2k-1)$-gon equation.}
 \end{cases}
 $
\end{enumerate}
\end{Proposition}

\begin{proof}
We only prove (1) for odd $n$. The others are proved in the same way.

Set $m=2k+1$. The partial composition $T^{(m)}\circ_lT'^{(n)}$ and tensor product $T^{(m)}\otimes T'^{(n)}$ are represented, respectively, by
 \begin{equation*}
 \adjustimage{valign=m}{partial_comp_and_tensor_prod.pdf}
 \end{equation*}
(The figure corresponds to the $m=7$ and $n=9$ case.) Then the proof for the partial composition case proceeds graphically as follows:
 \begin{equation*}
 \adjustimage{valign=M}{odd-gon_from_odd_odd_partial_comp.pdf}
 \end{equation*}
 The left-most and right-most figures are graphical representations of the $(m+n-3)$-gon equation for $T^{(m)}\circ_lT'^{(n)}$. The first equality follows from the commutativity of $T^{(m)}$ and $T'^{(n)}$, and the second from $T^{(m)}$ and $T'^{(n)}$ being solutions of the $m$- and $n$-gon equations, respectively.

 The tensor product case is more direct, as it is decomposed into three parts: the commutative part, $m$-gon equation, and $n$-gon equation
 \begin{equation*}
 \adjustimage{valign=M}{odd-gon_from_odd_odd_tensor_prod.pdf}\tag*{\qed}
 \end{equation*}\renewcommand{\qed}{}
\end{proof}

As a corollary, we are now able to prove all the conjectures proposed in \cite{muller2024structure}. Note that in~\cite{muller2024structure}, this corollary is stated in terms of the non-hatted version, and we have translated it here into our convention (see also Remark~\ref{rem:translation}).

\begin{Corollary}
 Here we consider set-theoretic polygon equations over a set X. Let $k$ be a positive integer.
 \begin{enumerate}\itemsep=0pt
 \item[$(1)$] Let $S^{(2k)}$ be a solution of the dual $2k$-gon equation. Then
 \begin{equation*}
 T^{(2k+1)}(a_1,\dots, a_k) \coloneqq \bigl(a_1,S^{(2k)}(a_1,\dots, a_k)\bigr)
 \end{equation*}
 is a solution of the $(2k+1)$-gon equation {\rm \cite[\emph{Conjecture}~7.32]{muller2024structure}}.
 \item[$(2)$] Let $S^{(2k)}$ be a solution of the dual $2k$-gon equation, and $u\in X$. Then
 \begin{equation*}
 T^{(2k+1)}(a_1,\dots, a_k) \coloneqq \bigl(u,S^{(2k)}(a_1,\dots, a_k)\bigr)
 \end{equation*}
 is a solution of the $(2k+1)$-gon equation if and only if $S^{(2k)}(u,\dots, u)=(u,\dots,u)$ {\rm \cite[\emph{Conjecture}~7.33]{muller2024structure}}.
 \item[$(3)$] Let $T^{(2k+1)}$ be a solution of the $(2k+1)$-gon equation. Then
 \begin{equation*}
 T^{(2k+2)}(a_1,\dots, a_k) \coloneqq \bigl(T^{(2k+1)}(a_1,\dots, a_k),a_k\bigr)
 \end{equation*}
 is a solution of the $(2k+2)$-gon equation {\rm \cite[\emph{Conjecture}~7.34]{muller2024structure}}.
 \item[$(4)$] Let $T^{(2k+1)}$ be a solution of the $(2k+1)$-gon equation, and $u\in X$. Then
 \begin{equation*}
 T^{(2k+2)}(a_1,\dots, a_k) \coloneqq \bigl(T^{(2k+1)}(a_1,\dots, a_k),u\bigr)
 \end{equation*}
 is a solution of the $(2k+2)$-gon equation if and only if $T^{(2k+1)}(u,\dots, u)=(u,\dots,u)$ {\rm \cite[\emph{Conjecture}~7.35]{muller2024structure}}.
 \item[$(5)$] Let $S^{(2k+1)}$ be a solution of the dual $(2k+1)$-gon equation. Then
 \begin{equation*}
 T^{(2k+2)}(a_1,\dots, a_k) \coloneqq \bigl(a_1,S^{(2k+1)}(a_1,\dots, a_k)\bigr)
 \end{equation*}
 is a solution of the $(2k+2)$-gon equation {\rm \cite[\emph{Conjecture}~7.36]{muller2024structure}}.
 \item[$(6)$] Let $S^{(2k+1)}$ be a solution of the dual $(2k+1)$-gon equation, and $u\in X$. Then
 \begin{equation*}
 T^{(2k+2)}(a_1,\dots, a_k) \coloneqq \bigl(u,S^{(2k+1)}(a_1,\dots, a_k)\bigr)
 \end{equation*}
 is a solution of the $(2k+2)$-gon equation if and only if $S^{(2k+1)}(u,\dots, u)=(u,\dots,u)$ {\rm \cite[\emph{Conjecture}~7.37]{muller2024structure}}.
 \end{enumerate}
\end{Corollary}

\begin{proof}
 We will only prove (1) and (2). The others can be proved similarly.

 (1) Define $\Delta\colon X\to X\times X$ by $\Delta(x) = (x, x)$. It is then easy to see that $\Delta$ is a solution of the 4-gon equation, and $\Delta \leftrightarrow S^{(2k)}$. Hence, \[
 T^{(2k+1)}=\Delta \circ_r S^{(2k)}
 \] is a solution of the $(2k+1)$-gon equation by Proposition~\ref{prop:commpolygon}\,(4).

 (2) For the ``if'' part, define $\Delta_u\colon X\to X\times X$ by $\Delta_u(x) = (u, x)$. It is also easy to see that~$\Delta_u$ is a solution of the 4-gon equation and $\Delta_u \leftrightarrow S^{(2k)}$. Hence, \[
 T^{(2k+1)}=\Delta_u \circ_r S^{(2k)}
 \] is a solution of the $(2k+1)$-gon equation by Proposition~\ref{prop:commpolygon}\,(4). For the ``only if'' part, substitute $(u,\dots,u)$ into the $(2k+1)$-gon equation. Then, by comparing the outputs at the position $a_1(=b_1)$, we obtain $S^{(2k)}(u,\dots, u)=(u,\dots,u)$.
\end{proof}

Note that the above corollary also works for the $n$-gon equation in a cartesian monoidal category, such as the category of vector spaces with the product given by the direct sum.

\section{Examples of solutions of polygon equations}\label{sec:examples_polygon_eq}

The following corollary follows directly from Proposition~\ref{prop:commpolygon} and will be useful when constructing examples of solutions of polygon equations.

\begin{Corollary}\label{cor:comm}\quad
 \begin{enumerate}\itemsep=0pt
 \item[$(1)$] Let $\{ T_i \}_{i=1,\cdots,n}$ be a family of solutions of the $5$-gon equation such that $T_i\leftrightarrow T_j$ for $i\neq j$, and let $\Delta$ be a solution of the $4$-gon equation such that $T_i\leftrightarrow\Delta$ for all $1\leq i \leq n$. Then $T_1 \circ_l T_2 \circ_l \cdots \circ_l T_n$ is a solution of the $(2n+3)$-gon equation, and $T_1 \circ_l T_2 \circ_l \cdots \circ_l T_n \circ_l \Delta$ is a solution of the $(2n+4)$-gon equation.
 \item[$(2)$] Let $\{ S_i \}_{i=1,\cdots,n}$ be a family of solutions of the dual $5$-gon equation such that $S_i\leftrightarrow S_j$ for $i\neq j$, and let $M$ be a solution of the dual $4$-gon equation such that $S_i\leftrightarrow M$ for all $1\leq i \leq n$. Then $S_1 \circ_r S_2 \circ_r \cdots \circ_r S_n$ is a solution of the dual $(2n+3)$-gon equation, and $S_1 \circ_r S_2 \circ_r \cdots \circ_r S_n \circ_r M$ is a solution of the dual $(2n+4)$-gon equation.
 \end{enumerate}
\end{Corollary}

\begin{Example}\label{ex:ex from comm Hopf}
 Let $H$ be a commutative and cocommutative bialgebra with product $M$ and coproduct $\Delta$.
 Then $\Delta$ is a solution of the 4-gon equation, and $M$ is a solution of the dual 4-gon equation.
 For an integer $n \geq 3$, we define $T^{(n)}$ by
 \begin{align*}
 T^{(2k+1)} \coloneqq \underbrace{\Delta \circ_r M \circ_l \Delta \circ_r M \circ_l \cdots \circ_{r} M}_{2k-2}, \qquad T^{(2k)} \coloneqq \underbrace{\Delta \circ_r M \circ_l \Delta \circ_r M \circ_l \cdots \circ_{l} \Delta}_{2k-3},
 \end{align*}
 and $T^{(3)} \coloneqq {\rm id}_H$.
 Let $T = \Delta \circ_r M$.
 From the bialgebra axioms and the (co)commutativity of~$M$ and $\Delta$, one can easily check that $T\leftrightarrow T$ and $T \leftrightarrow \Delta$.
 Since $T^{(2k+1)}=T\circ_l\cdots \circ_l T$ and $T^{(2k)}=T\circ_l\cdots \circ_l T\circ_l \Delta$, $T^{(n)}$ is a solution of the $n$-gon equation by Corollary~\ref{cor:comm}.

 Similarly, for an integer $n \geq 3$, if we define $S^{(n)}$ by
 \begin{align*}
 S^{(2k+1)} \coloneqq \underbrace{M \circ_l \Delta \circ_r M \circ_l \Delta \circ_r \cdots \circ_l \Delta}_{2k-2}, \qquad S^{(2k)} \coloneqq \underbrace{M \circ_l \Delta \circ_r M \circ_l \Delta \circ_r \cdots \circ_r M}_{2k-3},
 \end{align*}
 and $S^{(3)} \coloneqq {\rm id}_H$, then $S^{(n)}$ is a solution of the dual $n$-gon equation.
\end{Example}

\begin{Example}
 Let $(H_i,1_i,M_i,\varepsilon_i,\Delta_i)_{1\leq i\leq n}$ be a family of bialgebras with (co)unit.
 Set $H \coloneqq \bigotimes_{i=1}^{n}H_i$ and define pairs of product and coproduct $(\widehat{M}_i,\widehat{\Delta}_i)_{0\leq i\leq n-1}$ over $H$ by
 \begin{gather*}
 \widehat{M}_0((x_1 \otimes \cdots \otimes x_n) \otimes (y_1 \otimes \cdots \otimes y_n)) \coloneqq
 M_1(x_1 \otimes y_1) \otimes \cdots \otimes M_n(x_n \otimes y_n),\\
 \widehat{\Delta}_0(x_1 \otimes \cdots \otimes x_n) \coloneqq (x_{1(1)_1} \otimes \cdots \otimes x_{n(1)_n}) \otimes (x_{1(2)_1} \otimes \cdots \otimes x_{n(2)_n}),
 \end{gather*}
 where we use Sweedler notation $\Delta_i(x)=x_{(1)_i}\otimes x_{(2)_i}$, and for $1 \leq i \leq n-1$,
 \begin{gather*}
 \widehat{M}_i((x_1 \otimes \cdots \otimes x_n) \otimes (y_1 \otimes \cdots \otimes y_n)) \\
 \qquad{} \coloneqq
 \prod_{j=1}^{i}\varepsilon_j(y_j)\prod_{j=i+1}^{n}\varepsilon_j(x_j)(x_1 \otimes \cdots \otimes x_i \otimes y_{i+1} \otimes \cdots \otimes y_n),\\
 \widehat{\Delta}_i(x_1 \otimes \cdots \otimes x_n) \coloneqq (x_1 \otimes \cdots \otimes x_i \otimes 1_{i+1} \otimes \cdots \otimes 1_{n}) \otimes (1_{1} \otimes \cdots \otimes 1_{i} \otimes x_{i+1} \otimes \cdots \otimes x_n).
 \end{gather*}
 Then the tuple $(H, \widehat{M}_i,\widehat{\Delta}_i)$ forms a bialgebra for each $i$.
 If we set $T_i\coloneqq\widehat{\Delta}_i \circ_r \widehat{M}_i$, a tedious but straightforward calculation shows that $T_i\leftrightarrow T_j$ for $i\neq j$ and $T_i\leftrightarrow \widehat{\Delta}_{j+1}$ for $0\leq i \leq j$.
 Thus, by Corollary \ref{cor:comm},
 \begin{gather*}
 T^{(2k+1)} \coloneqq \widehat{\Delta}_1 \circ_r \widehat{M}_1 \circ_l \widehat{\Delta}_2 \circ_r \widehat{M}_2 \circ_l \cdots \circ_{r} \widehat{M}_{k-1}, \\
 T^{(2k+2)} \coloneqq \widehat{\Delta}_1 \circ_r \widehat{M}_1 \circ_l \widehat{\Delta}_2 \circ_r \widehat{M}_2 \circ_l \cdots \circ_{l} \widehat{\Delta}_{k-1}
 \end{gather*}
 are solutions of the $(2k+1)$- and $(2k+2)$-gon equations, respectively.
\end{Example}

\begin{Example}\label{ex:polygon from n-groupid}
 A strict $n$-category $\mathcal{C}$ consists of sets $(\mathcal{C}_0,\mathcal{C}_1,\cdots,\mathcal{C}_n)$ and maps
 \begin{gather*}
 s_k, t_k \colon \ \mathcal{C}_k \to \mathcal{C}_{k-1} \qquad \text{for } 1 \leq k \leq n ,\\
 1_k \colon \ \mathcal{C}_k \to \mathcal{C}_{k+1} \qquad \text{for } 0 \leq k \leq n-1 ,
\\
 \circ_{p,q} \colon \ \{(f,g)\in \mathcal{C}_p \times \mathcal{C}_p\mid s_{pq}(f)=t_{pq}(g)\} \to \mathcal{C}_p \qquad \text{for } 0 \leq q < p,
 \end{gather*}
 where $s_{pq} \coloneqq s_{q+1}\cdots s_{p-1}s_p$, $t_{pq} \coloneqq t_{q+1}\cdots t_{p-1}t_p$ and $1_{qp}=1_{p-1}\cdots 1_{q+1}1_{q}$.
 Elements of $\mathcal{C}_p$ are called $p$-morphisms.
 We will write $\circ_q$ for $\circ_{p,q}$ when $p$ is obvious from the context.

 These maps should satisfy the following axioms:
 \begin{itemize}\itemsep=0pt
 \item $s_{p-1}s_{p}=s_{p-1}t_p$, $t_{p-1}s_p=t_{p-1}t_p$ and $s_{p+1}1_p={\rm id}_{\mathcal{C}_p}=t_{p+1}1_p$.
 \item For $f,g\in \mathcal{C}_p$ and $q<p$ with $s_{pq}(f)=t_{pq}(g)$,
 \begin{align*}
 & s_{p}(f\circ_q g)=
 \begin{cases}
 s_p(f)\circ_q s_p(g) &\text{if $q<p-1$,} \\
 s_p(g) &\text{if $q=p-1$,}
 \end{cases} \\
 & t_{p}(f\circ_q g)=
 \begin{cases}
 t_p(f)\circ_q t_p(g) &\text{if $q<p-1$,} \\
 t_p(f) &\text{if $q=p-1$.}
 \end{cases}
 \end{align*}
 \item For $f, g ,h\in \mathcal{C}_p$, $(f\circ_q g)\circ_q h=f\circ_q (g\circ_q h)$.
 \item For $q<p<r$ and $f, g, h, k\in\mathcal{C}_r$ with $t_q(f)=s_q(g)$, $t_q(h)=s_q(k)$ and $t_p(f)=s_p(h)$,
 \begin{align*}
 (f\circ_q g)\circ_p (h\circ_q k)=(f\circ_p h)\circ_q(g\circ_p k).
 \end{align*}
 \item For $f\in\mathcal{C}_p$ and $q<p$, $f\circ_q 1_{qp}(s_{pq}(f))=f=1_{qp}(t_{pq}(f))\circ_q f$.
 \end{itemize}

 A strict $n$-groupoid $\mathcal{G}=(\mathcal{G}_0,\mathcal{G}_1,\dots,\mathcal{G}_n)$ is a strict $n$-category where each $i$-morphism is invertible with respect to $\circ_j$ for $0\leq j<i$, i.e., for any $1\leq i \leq n$, any $0\leq j <i$ and any $a\in \mathcal{G}_i$, there exist $b\in\mathcal{G}_i$ such that $s_{ij}(a)=t_{ij}(b)$, $s_{ij}(b)=t_{ij}(a)$, $a \circ_{j} b=1_{ji}(t_{ij}(a))$, and $b \circ_{j} a=1_{ji}(s_{ij}(a))$.

 Let $H$ be the vector space spanned by all the $n$-morphisms of the strict $n$-groupoid $\mathcal{G}$.
 For each $0 \leq i \leq n-1$, set
 \begin{equation*}
 \Delta_i \colon \ H \to H \otimes H,\qquad f\mapsto f \otimes f, \qquad M_i \colon \ H \otimes H \to H, \qquad f \otimes g\mapsto \delta^{s_{ni}(f)}_{t_{ni}(g)}f\circ_{i}g,
 \end{equation*}
 where \smash{$\delta^{s_{ij}(f)}_{t_{ij}(g)}=1$ if $s_{ij}(f)=t_{ij}(g)$} and $0$ otherwise.
 Then for any $i$, the tuple $(H, \Delta_i ,M_i)$
 forms a (weak) bialgebra.
 For $2\leq k \leq \frac{n-2}{2}$, if we set
 \begin{gather*}
 T^{(2k+1)} \coloneqq \Delta_0 \circ_r M_0 \circ_l \Delta_1 \circ_r \cdots \circ_{r} M_{k-2}, \qquad T^{(2k)} \coloneqq \Delta_0 \circ_r M_0 \circ_l \Delta_1 \circ_l \cdots \circ_{l} \Delta_{k-2},\\
 S^{(2k+1)} \coloneqq M_{n-1} \circ_l \Delta_{n-1} \circ_r \cdots \circ_l \Delta_{n-k+1}, \qquad S^{(2k)} \coloneqq M_{n-1} \circ_l \Delta_{n-1} \circ_r \cdots \circ_r M_{n-k+1},
 \end{gather*}
 then an argument similar to those above shows that these are solutions of the polygon and dual polygon equations.
\end{Example}

\begin{Remark}
 These examples suggest the following algebraic structure for constructing solutions of polygon equations: Let $(H, \mu_1, \dots ,\mu_n)$ be a tuple where $H$ is a vector space, and $\mu_i$ is an associative product when $i$ is even, and a coassociative coproduct when $i$ is odd.
 Furthermore, $\mu_{i+1}$ should be commutative with $\mu_1\circ_r\mu_2\circ_l\mu_3\circ_r\mu_4\circ_l\cdots\circ_{\text{($r$ or $l$)}}\mu_i$ for each $i$.
 In other words, the map $\mu_1\circ_r\mu_2\circ_l\mu_3\circ_r\mu_4\circ_l\cdots\circ_{\text{($r$ or $l$)}}\mu_i$ should be an algebra or a coalgebra map with respect to $\mu_{i+1}$. Then, by Proposition~\ref{prop:commpolygon}, the map $\mu_1\circ_r\mu_2\circ_l\cdots\circ_{\text{($r$ or $l$)}}\mu_n$ solves the $(n+3)$-gon equation.
 For $n=1$ and $n=2$, this matches the definitions of a coalgebra and a bialgebra (without (co)unit), respectively, and for $n=3$, this seems to be related to the (dual) notion of a trialgebra \cite{crane1994four, grosse2000trialgebraic, grosse2000second, pfeiffer20072}.
\end{Remark}

\section{Solutions of simplex equations from polygon equations}\label{sec:simplex_from_polygon}

In this section, we explore the relationship between simplex equations and polygon equations. The main theorems (Theorems~\ref{thm:fromnton-1} and \ref{thm:fromnton-2}) generalize the result from~\cite{kashaev1998pentagon}.

We first define simplex equations.
\begin{Definition}[simplex equation] For a positive integer $n$, the $n$-simplex equation for a linear map $R\colon V^{\otimes n}\to V^{\otimes n}$ is an equation in $\operatorname{End}\bigl(V^{\otimes n(n+1)/2}\bigr)$ of the form
 \begin{equation*}
 R_{a_1}R_{a_2}\cdots R_{a_{n+1}} = R_{a_{n+1}}R_{a_n}\cdots R_{a_1},
 \end{equation*}
 where $[ a_1, a_2, \dots, a_{n+1}]^{\mathsf{T}}$ is the matrix $A_{2n+1}$ defined in Definition~\ref{def:oddgon}.

 Alternatively, the $n$-simplex equation can be defined graphically by
 \begin{equation*}
 \includegraphics[scale=1]{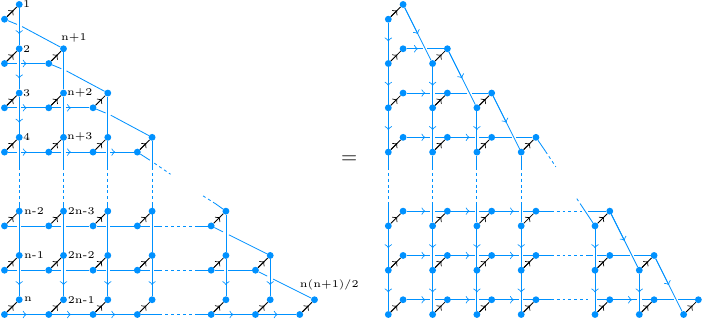}
 \end{equation*}
 (the left-hand side of the figure is read from bottom to top, and the right-hand side is read from top to bottom).
\end{Definition}

\begin{Example}
 The $n$-simplex equations for lower $n$ are given as follows:
\begin{align*}
 &\text{1-simplex:}\quad R_{1}R_{1}=R_{1}R_{1},\\
 &\text{2-simplex:}\quad R_{12}R_{13}R_{23}=R_{23}R_{13}R_{12},\\
 &\text{3-simplex:}\quad R_{123}R_{145}R_{246}R_{356}=R_{356}R_{246}R_{145}R_{123},\\
 &\text{4-simplex:}\quad R_{1234}R_{1567}R_{2589}R_{368,10}R_{479,10}=R_{479,10}R_{368,10}R_{2589}R_{1567}R_{1234}.
\end{align*}
\end{Example}

We have the following property of solutions of simplex equations:

\begin{Proposition}
\label{prop:partialtracesimplex}
 If $R$ is an invertible solution of the $n$-simplex equation over a finite-di\-men\-sional vector space $V$, then $\operatorname{tr}_l(R)$ and $\operatorname{tr}_r(R)$ are solutions of the $(n-1)$-simplex equation.
\end{Proposition}

\begin{proof}
 The proof proceeds as in Proposition~\ref{prop:usefulpolygon}\,(5). For the $\operatorname{tr}_l(R)$ case, compose with $R_{a_1}^{-1}$ and take the trace over the position $a_1$ on both sides. For the $\operatorname{tr}_r(R)$ case, replace~$a_1$ with~$a_{n+1}$.
\end{proof}

\subsection{Mixed relation}\label{sec:mixed_relation}

Let us define the mixed relation explicitly, which corresponds to the equation associated with the mixed order in \cite{dimakis2015simplex}.

\begin{Definition}[mixed relation for odd-gon]\label{def:mixed_odd}
 Let $T$ be a solution of $(2k+1)$-gon equation and~$S$ a~solution of the dual $(2k+1)$-gon equation. Then the mixed relation is
 \begin{equation*}
 T_{d_{k+1}}S_{e_k}T_{d_k}\cdots S_{e_1}T_{d_{1}} = S_{f_1}T_{g_1}S_{f_2}\cdots T_{g_k}S_{f_{k+1}},
 \end{equation*}
 where the multi-indices $d_i$, $e_i$, $f_i$ and $g_i$ are as follows:
 \begin{align*}
& E_{2k+1}= \begin{bmatrix}
 e_1\\
 \vdots\\
 e_k
 \end{bmatrix} = \begin{bNiceArray}{c|cccc}[margin]
 1 & k+1 & k+2 & \cdots & 2k-1 \\\hline
 2 & \Block{4-4}{E_{2k-1}+[2k-1]_{k-1,k-1}} \\
 3\\
 \vdots \\
 k \\
 \end{bNiceArray},\\
 & E_3 = [ 1 ],\\
& G_{2k+1}= \begin{bmatrix}
 g_1\\
 \vdots\\
 g_k
 \end{bmatrix} = (E_{2k+1})^{\mathsf{T}},\\
& D_{2k+1}= \begin{bmatrix}
 d_1\\
 \vdots\\
 d_{k+1}
 \end{bmatrix} = \begin{bNiceArray}{c|w{c}{0.5cm}w{c}{0.5cm}w{c}{0.5cm}w{c}{0.5cm}}[margin]
 1 & 2 & 3 & \cdots & k \\\hline
 1 & \Block{5-4}{D_{2k-1}+[2k-1]_{k,k-1}} \\
 k+1\\
 k+2\\
 \vdots \\
 2k-1 \\
 \end{bNiceArray},\\
 & D_3 = \left[\!\begin{array}{c}
 1\\
 1\\
 \end{array}\!\right],\\
 &F_{2k+1}= \begin{bmatrix}
 f_1\\
 \vdots\\
 f_{k+1}
 \end{bmatrix} = \begin{bNiceArray}{c|cccc}[margin]
 1 & k+1 & k+2 & \cdots & 2k-1 \\\hline
 1 & \Block{4-4}{F_{2k-1}+[2k-1]_{k,k-1}}\\
 2\\
 \vdots \\
 k \\
 \end{bNiceArray},\\
 & F_3 = \left[\!\begin{array}{c}
 1\\
 1\\
 \end{array}\!\right].
 \end{align*}

 The graphical definition is
 \begin{equation*}
 \includegraphics[scale=1]{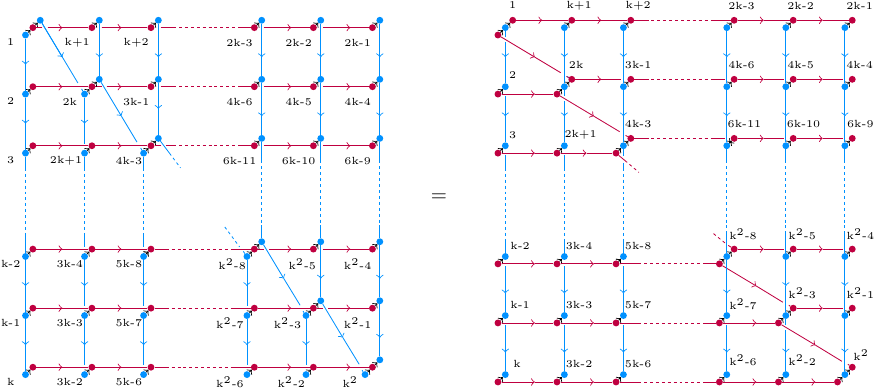}
 \end{equation*}
 (the left-hand side is read from bottom to top, and the right-hand side is read from top to bottom).
\end{Definition}

For $k=2$, this recovers the ten-term relation of the (dual) pentagon equations in \cite{kashaev1998pentagon}.

\begin{Definition}[mixed relation for even-gon]\label{definition6.5}
 Let $T$ be a solution of the $2k$-gon equation and~$S$ a solution of the dual $2k$-gon equation. Then the mixed relation is graphically defined by
 \begin{equation*}
 \includegraphics[scale=1]{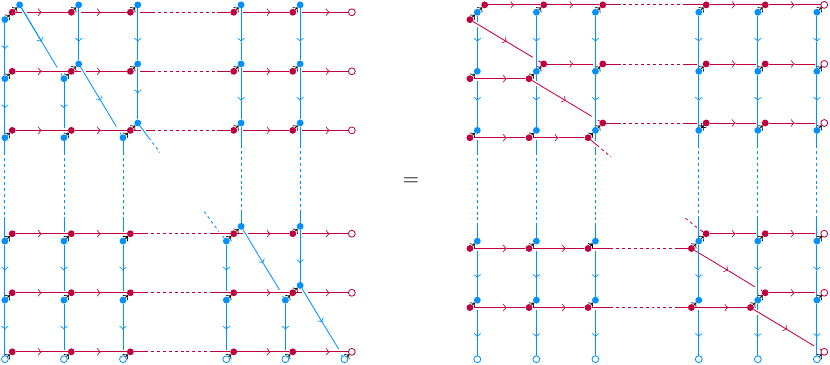}
 \end{equation*}
 (the left-hand side is read from bottom to top, and the right-hand side is read from top to bottom).\footnote{Writing the mixed relation for even-gon in our tensor notation is complicated, and so we define it graphically.}
\end{Definition}

\begin{Proposition}
 Let $T$ be an invertible solution of the $(2k+1)$-gon equation, and $S$ an invertible solution of the dual $(2k+1)$-gon equation. If $(T,S)$ satisfies the mixed relation, then the pairs
 $\bigl(\operatorname{dim}(V)^{-1}\operatorname{tr}_l(T),\allowbreak\operatorname{dim}(V)^{-1}\operatorname{tr}_l(S)\bigr)$ and $\bigl(\operatorname{dim}(V)^{-1}\operatorname{tr}_r(T), \operatorname{dim}(V)^{-1}\operatorname{tr}_r(S)\bigr)$ do as well.
\end{Proposition}

\begin{proof}
 By Proposition \ref{prop:usefulpolygon}\,(5), $\operatorname{dim}(V)^{-1}\operatorname{tr}_l(T)$ is a solution of the $(2k-1)$-gon equation (and similarly, $\operatorname{dim}(V)^{-1}\operatorname{tr}_l(S)$ is a solution of the dual $(2k-1)$-gon equation).

 The figure shows the graphical proof for the 7-gon case (the general case follows in the exact same way):
 \begin{equation*}
 \adjustimage{valign=M}{mixed_trace.pdf}
 \end{equation*}
 (inverses are shown in corresponding colored lines with empty dots). Composing with $S_{e_1}^{-1}$ $\bigl(=S_{f_1}^{-1}\bigr)$ and $T_{d_1}^{-1}$ $\bigl(=T_{g_1}^{-1}\bigr)$, and taking the partial trace over the positions $e_1$ $(=f_1)$ and $d_1$ $(=g_1)$ shows that $\bigl(\operatorname{dim}(V)^{-1}\operatorname{tr}_l(T), \operatorname{dim}(V)^{-1}\operatorname{tr}_l(S)\bigr)$ satisfies the mixed relation.
\end{proof}

\subsection[From n-gon to (n-1)-simplex]{From $\boldsymbol{n}$-gon to $\boldsymbol{(n-1)}$-simplex}
We construct a solution of the $(n-1)$-simplex equation from a pair consisting of a solution of the $n$-gon and a solution of the dual $n$-gon equation, satisfying the mixed relation. Although this result was previously discussed in \cite{dimakis2015simplex} and proved in \cite{dimakis2021grassmannian} for the odd-gon equations, we provide a~more direct proof using graphical calculus here.

\begin{Theorem}\label{thm:fromnton-1}
 Let $T$ be a solution of the $n$-gon equation, and $S$ be a solution of its dual. Suppose that $T$ and $S$ satisfy the mixed relation. Then
 \begin{align*}
 R^{(n-1)} &\coloneqq \begin{cases}
 \sigma_{1,2}\sigma_{3,4}\cdots \sigma_{2k-1,2k}T_{2,4,\dots,2k}S_{1,3,\dots,2k-1} & \text{if $ n=2k+1$,}\\
 \sigma_{1,2}\sigma_{3,4}\cdots \sigma_{2k-3,2k-2}T_{2,4,\dots,2k-2}S_{1,3,\dots,2k} & \text{if $n=2k$,}
 \end{cases}\\
 \intertext{or graphically,}
 R^{(n-1)} &\coloneqq \begin{cases}
 \adjustimage{valign=M}{n-1_from_n_odd.pdf} & \text{if $ n=2k+1$,}\\[15pt]
 \adjustimage{valign=M}{n-1_from_n_even.pdf} & \text{if $n=2k$}
 \end{cases}
 \end{align*}
 is a solution of the $(n-1)$-simplex equation, where $\sigma_{i,j}$ flips the $i$-th and $j$-th factors if $i\ne j$ and is the identity if $i=j$, and is represented in green.
\end{Theorem}

\begin{proof}
 We prove the $2k$-gon case. The odd-gon case is proved similarly, or the reader may also refer to \cite[Section~2.5]{dimakis2021grassmannian}.

 We first substitute a map of the form
 \begin{equation*}
 \sigma_{1,2}\sigma_{3,4}\cdots \sigma_{2k-3,2k-2}F\; (= \adjustimage{valign=b}{mapF2n-1_even.pdf}\quad\text{ graphically}),
 \end{equation*}
 where $F\colon V^{\otimes (2k-1)}\to V^{\otimes (2k-1)}$ is any map, into the $(2k-1)$-simplex equation and separate the transpositions $\sigma_{i,j}$ from $F$'s as follows:
 \begin{align*}
 &\text{(LHS)} = \quad \adjustimage{valign=m,scale=0.61}{2n-1-simplex_from_even_lhs1.pdf} \!=\quad \adjustimage{valign=m,scale=0.61}{2n-1-simplex_from_even_perm_term.pdf} \circ\quad \underbrace{\adjustimage{valign=m,scale=0.61}{2n-1-simplex_from_even_lhs2.pdf}}_{\text{(I)}}\\
 &\text{(RHS)} = \quad \adjustimage{valign=m,scale=0.61}{2n-1-simplex_from_even_rhs1.pdf} \!=\quad \adjustimage{valign=m,scale=0.61}{2n-1-simplex_from_even_perm_term.pdf} \circ\quad \underbrace{\adjustimage{valign=m,scale=0.61}{2n-1-simplex_from_even_rhs2.pdf}}_{\text{(II)}}
 \end{align*}
 Note that we used the following formulas for transpositions:
 \begin{gather*}
 \adjustimage{valign=m}{permutation_prop1.pdf} \qquad
 \adjustimage{valign=m}{permutation_prop2.pdf} \qquad \adjustimage{valign=m}{permutation_prop3.pdf} \end{gather*}
 It now suffices to verify the equality $\text{(I)}=\text{(II)}$ for the map
 \begin{equation*}
 F =\; \adjustimage{valign=m}{2n-1map_even.pdf} \end{equation*}
 This is also done by graphical calculus:
 \begin{equation*}
 \text{(I)} = \ \adjustimage{valign=m, scale=0.75}{2n-1-simplex_from_even_lhs3.pdf} \qquad \text{(II)} = \ \adjustimage{valign=m, scale=0.75}{2n-1-simplex_from_even_rhs3.pdf}
 \end{equation*}
 Let us assign coordinates of the input/output vector spaces as in Figure~\ref{fig:coordinates of input/output}.
 Then, one can see that the diagrams $\text{(I)}$ and $\text{(II)}$ decompose into three independent parts (indeed, this decomposition is called the three color decomposition in \cite{dimakis2015simplex}):
 \begin{itemize}\itemsep=0pt
 \item the part lying on input coordinates of type $(\text{even},\text{even})$, forming the $2k$-gon equation,
 \item the part lying on input coordinates of type $(\text{odd},\text{odd})$, forming the dual $2k$-gon equation,
 \item the part lying on input coordinates of type $(\text{even},\text{odd})$ or $(\text{odd},\text{even})$, forming the mixed relation.
 \end{itemize}
 Note that the mixed relation appearing in the diagrams $\text{(I)}$ and $\text{(II)}$ is ``folded'' along the diagonal line, and if we unfold it by moving each $(\text{odd},\text{even})$ coordinate $(a,b)$ to $(b,a)$, this matches the graphical definition of the mixed relation given in Definition~\ref{definition6.5}.
 \begin{figure}[!ht]
 \centering
 \includegraphics[width=0.55\linewidth]{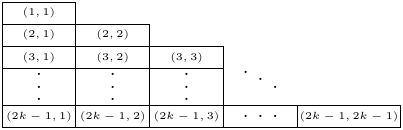}
 \caption{Coordinates of the input/output vector spaces.}
 \label{fig:coordinates of input/output}
 \end{figure}
\end{proof}

\subsection[From n-gon to (n-2)-simplex]{From $\boldsymbol{n}$-gon to $\boldsymbol{(n-2)}$-simplex}
The same condition as in Theorem~\ref{thm:fromnton-1} also gives rise to a solution of the $(n-2)$-simplex equation. This generalizes the result from \cite[Propositions~4, 5]{kashaev1998pentagon}.

\begin{Theorem}\label{thm:fromnton-2}
 Let $T$ be a solution of the $n$-gon equation and $S$ be a solution of its dual. Suppose that $T$ and $S$ satisfy the mixed relation. Then
 \begin{equation*}
 R^{(n-2)} \coloneqq \begin{cases}
 \adjustimage{valign=M}{n-2_from_n_odd.pdf} & \text{if $ n=2k+1$,}\\[15pt]
 \adjustimage{valign=M}{n-2_from_n_even.pdf} & \text{if $n=2k$}
 \end{cases}
 \end{equation*}
 is a solution of the $(n-2)$-simplex equation.\footnote{Writing the map $R^{(n-2)}$ in our tensor notation is complicated, and so we represent it graphically.}
\end{Theorem}

\begin{proof}
 We prove the $(2k+1)$-gon case. The even-gon case is proved similarly.

 We take a similar step to the proof of Theorem~\ref{thm:fromnton-1}. We substitute a map of the form
 \begin{equation*}
 \sigma_{2,3}\sigma_{4,5}\cdots\sigma_{2k-2,2k-1}F\ (= \adjustimage{valign=b}{mapF2n-1_odd.pdf}\quad\text{ graphically}),
 \end{equation*}
 where $F\colon V^{\otimes (2k-1)}\to V^{\otimes (2k-1)}$ is any map, into the $(2k-1)$-simplex equation and separate transpositions from $F$'s as follows:
 \begin{align*}
 &\text{(LHS)} = \, \adjustimage{valign=m,scale=0.660}{2n-1-simplex_from_odd_lhs1.pdf} \!=\ \adjustimage{valign=M,scale=0.660, raise=-6pt}{2n-1-simplex_from_odd_perm_term.pdf} \circ\quad \underbrace{\adjustimage{valign=m,scale=0.660, raise=-2pt}{2n-1-simplex_from_odd_lhs2.pdf}}_{\text{(I)}} \\
 &\text{(RHS)} = \, \adjustimage{valign=m,scale=0.660}{2n-1-simplex_from_odd_rhs1.pdf} \!=\ \adjustimage{valign=M,scale=0.660, raise=-3pt}{2n-1-simplex_from_odd_perm_term.pdf} \circ\quad \underbrace{\adjustimage{valign=m,scale=0.660}{2n-1-simplex_from_odd_rhs2.pdf}}_{\text{(II)}}
 \end{align*}
 It now suffices to verify the equality $\text{(I)}=\text{(II)}$ for the map
 \begin{equation*}
 F = \adjustimage{valign=m}{2n-1map_odd.pdf} \end{equation*}
 This is also done by graphical calculus. However, unlike Theorem~\ref{thm:fromnton-1}, the diagrams are not decomposed into three independent parts corresponding to the $(2k+1)$-gon, its dual, and the mixed relation. Instead, these equations are applied in turn. We have
 \begin{align*}
 \text{(I)} &= \ \adjustimage{valign=m, scale=0.745}{2n-1-simplex_from_odd1.pdf}\\[10pt]
 &= \ \adjustimage{valign=M, scale=0.745}{2n-1-simplex_from_odd2.pdf} \\[10pt]
 &= \ \adjustimage{valign=M, scale=0.745}{2n-1-simplex_from_odd3.pdf} \\[10pt]
 &= \ \adjustimage{valign=M, scale=0.745}{2n-1-simplex_from_odd4.pdf} \\[10pt]
 &= \ \adjustimage{valign=M, scale=0.745}{2n-1-simplex_from_odd5.pdf} \\[10pt]
 &= \ \adjustimage{valign=M, scale=0.745}{2n-1-simplex_from_odd6.pdf} = \text{(II)}.
 \end{align*}
 As in the proof of Theorem~\ref{thm:fromnton-1}, we put coordinates of the input/output vector spaces as in Figure~\ref{fig:coordinates of input/output}. Then,
 \begin{itemize}\itemsep=0pt
 \item the third equality follows from the dual $(2k+1)$-gon equation, consisting of the spaces with coordinates of type $(1,1)$, $(\text{even},1)$, or $(\text{even},\text{even})$,
 \item the fourth equality follows from the mixed relation, consisting of the spaces with coordinates of type $(1,*)$, $(\text{even},\text{odd})$, or $(\text{odd},\text{even})$ (again the mixed relation is ``folded'' just as in the case of the proof of Theorem~\ref{thm:fromnton-1}), and
 \item the fifth equality follows from the $(2k+1)$-gon equation, consisting of the spaces with coordinates of type $(\text{odd},\text{odd})$. \hfill \qed
 \end{itemize}\renewcommand{\qed}{}
\end{proof}

\begin{figure}[!ht]
 \centering
 \includegraphics[scale=1]{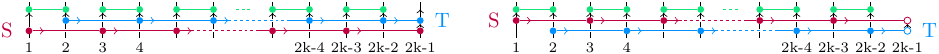}
 \caption{Right partial trace version of $R^{(n-2)}$, from odd-gon (left), and even-gon (right).}
 \label{fig:from_n_to_n-2(right_trace)}
\end{figure}

\begin{Remark}
 One may notice that $R^{(n-2)}=\operatorname{tr}_l\bigl(R^{(n-1)}\bigr)$, i.e., $R^{(n-2)}$ is the left partial trace of $R^{(n-1)}$ in Theorem~\ref{thm:fromnton-1}. Thus, if $R^{(n-1)}$ is invertible, Theorem~\ref{thm:fromnton-2} follows from Proposition~\ref{prop:partialtracesimplex}. However, even if $R^{(n-1)}$ is not invertible, the theorem still holds.

 Similarly, one can show that the right partial trace of $R^{(n-1)}$ also satisfies the $(n-2)$-simplex equation (Figure~\ref{fig:from_n_to_n-2(right_trace)}).
\end{Remark}

\begin{Remark}
 Note that our proofs of Theorems~\ref{thm:fromnton-1} and \ref{thm:fromnton-2} also work in the non-constant setting, in which each $T$ and $S$ is appropriately parametrized by an $(n-2)$-face of the simplex~$\Delta^{n-1}$, see Appendix~\ref{sec:nonconstantthm}.
\end{Remark}

\subsection{Examples}

In \cite{dimakis2021grassmannian}, the authors constructed a solution of the (non-constant) $2n$-simplex equation from solutions of the (non-constant) $(2n+1)$-gon and dual $(2n+1)$-gon equations, which form a pair satisfying the mixed relation. Using Theorem \ref{thm:fromnton-2}, we know that their construction further descends to a solution of the $(2n-1)$-simplex equation. Another example of a pair of solutions of the 5-gon and dual 5-gon equations satisfying the mixed relation is given in \cite{evripidou2024quadrirational}.

Let us give a few examples of pairs of solutions of the $n$-gon and dual $n$-gon equations satisfying the mixed relation.

\begin{Example}\label{ex:tenterm from Hopf}
 We apply the idea from \cite[Section~4]{kashaev1998pentagon} to construct solutions of the 3- and 4-simplex equations from solutions of the pentagon equation and its dual satisfying the mixed relation.

 Let $H=(H,1,M,\varepsilon,\Delta,s)$ be a finite-dimensional Hopf algebra. The $O$-double $O(H)$ of $H$ is an algebra $H^*\otimes H\otimes H^*$ with unit $\varepsilon\otimes 1\otimes\varepsilon$ and multiplication
 \begin{equation*}
 (f\otimes x\otimes \varphi)(g\otimes y\otimes\psi) = f(x_{(1)}\rightharpoonup g)\otimes x_{(2)}(\varphi_{(1)}\rightharpoonup y)\otimes\varphi_{(2)}\psi
 \end{equation*}
for $f,g,\varphi,\psi\in H^*$, $x,y\in H$, where
\begin{equation*}
 \langle x\rightharpoonup f,y\rangle=f(yx),\qquad f\rightharpoonup x=x_{(1)}f(x_{(2)}),\qquad f\in H^*,\quad x,y\in H,
 \end{equation*}
 are the left actions of $H$ on $H^*$, and of $H^*$ on $H$, respectively~\cite{novikov2000various}.\footnote{The subalgebra $H^*\otimes H\otimes \varepsilon$ is isomorphic to the Heisenberg double of $H$.} For an arbitrary basis $\{e_i\}_i$ of $H$ and its dual basis $\{e^i\}_i$, Kashaev and Sergeev showed that
 \begin{gather*}
 \mathsf{T}=(\varepsilon\otimes e_i\otimes\varepsilon)\otimes(e^i\otimes 1\otimes\varepsilon)\!\qquad\text{and}\!\qquad \mathsf{S}=(\varepsilon\otimes e_i\otimes\varepsilon)\otimes(\varepsilon\otimes 1\otimes e^i)\in O(H)\otimes O(H)
 \end{gather*}
 satisfy the pentagon equation and its dual, respectively, and moreover satisfy the mixed relation (in particular, \eqref{eq:1}--\eqref{eq:3} below).

 We now consider the action $F\colon O(H)\to\operatorname{End}(H^*)$ of $O(H)$ on $H^*$ given by\footnote{The restriction of $F$ to the subalgebra $H^*\otimes H\otimes \varepsilon$ is known as the Fock representation.}
 \begin{equation*}
 F(f\otimes x\otimes\varphi)(\alpha)=f(x\rightharpoonup\alpha)s^{*-1}(\varphi)\quad (\alpha\in H^*).
 \end{equation*}
 We extend this diagonally to the action $F\otimes F\colon O(H)\otimes O(H)\to\operatorname{End}(H^*\otimes H^*)$ and see the images of $\mathsf{T}$ and $\mathsf{S}$.
 We have for $\alpha,\beta\in H^*$
 \begin{align*}
 (F\otimes F(\mathsf{T}))(\alpha\otimes\beta) = \alpha_{(1)}\otimes\alpha_{(2)}\beta \qquad\text{and}\qquad (F\otimes F(\mathsf{S}))(\alpha\otimes\beta) = \alpha_{(1)}\otimes\beta s^{*-1}(\alpha_{(2)}).
 \end{align*}
 Since $\mathsf{T}$ and $\mathsf{S}$ satisfy the pentagon equation and its dual, respectively, and also the mixed relation, the same properties descend to $T\coloneqq F\otimes F(\mathsf{T})$ and $S\coloneqq F\otimes F(\mathsf{S})$ (of course, one can verify this directly). If we start this argument with $H^*$, we eventually arrive at the two maps $T,S\colon H\otimes H\to H\otimes H$:
 \begin{equation*}
 T(x\otimes y) = x_{(1)}\otimes x_{(2)}y,\qquad S(x\otimes y) = x_{(1)}\otimes ys^{-1}(x_{(2)}).
 \end{equation*}
 The solutions of the 3- and 4-simplex equations obtained from them are
 \begin{gather*}
 R^{(3)}(x\otimes y\otimes z) = x_{(1)}\otimes zs^{-1}(x_{(3)})\otimes x_{(2)}y,\\ R^{(4)}(x\otimes y\otimes z\otimes w) = y_{(1)}\otimes x_{(1)}\otimes y_{(2)}w\otimes zs^{-1}(x_{(2)}),
 \end{gather*}
 respectively.

 \begin{Remark}
 Additionally, one can see that $S\circ T\colon H\otimes H\to H\otimes H$ satisfies the Yang--Baxter (2-simplex) equation.\footnote{We have $S\circ T(x\otimes y)=x_{(1)}\otimes x_{(3)}ys^{-1}(x_{(2)})$, and this map, post-composed with the transposition $\sigma$, coincides with the solution $T'$ of the braid equation given in \cite{woronowicz1991solutions}.} Kashaev~\cite{kashaev1995heisenberg} also showed that $S_{14}T_{13}S_{24}T_{23}\colon(H\otimes H)\otimes(H\otimes H)\to (H\otimes H)\otimes(H\otimes H)$ solves the Yang--Baxter equation. A natural question is whether these constructions extend to the general case, that is, whether one can construct a solution of the $(n-3)$-simplex equation from solutions of the $n$-gon and dual $n$-gon equations. We leave this question for future work.
 \end{Remark}
\end{Example}

Let $T$ and $S$ be solutions of the pentagon and dual pentagon equations, respectively. Kashaev and Sergeev~\cite[equation(4.1)]{kashaev1998pentagon} showed that if $T$ and $S$ satisfy
\begin{align}
& T_{13}S_{23}= S_{23}T_{13},\label{eq:1}\\
& T_{23}T_{13}S_{12}= S_{12}T_{23},\label{eq:2}\\
& T_{12}S_{13}S_{23}= S_{23}T_{12},\label{eq:3}
\end{align}
then they satisfy the mixed relation. The graphical representations of these relations are
\begin{equation*}
 \adjustimage{valign=m}{eq1-1.pdf}\ = \ \adjustimage{valign=m}{eq1-2.pdf}\,, \qquad\adjustimage{valign=m}{eq2-1.pdf}\ = \ \adjustimage{valign=m}{eq2-2.pdf}\,, \qquad\adjustimage{valign=m}{eq3-1.pdf}\ = \ \adjustimage{valign=m}{eq3-2.pdf}\,,
\end{equation*}
respectively (the blue line represents $T$, and the red line $S$ ).

We now look for a similar condition for solutions of higher-order polygon equations. In particular, we consider solutions of the form $T^{(2k+1)}\coloneqq T\circ_lT\circ\cdots\circ_lT$ and $S^{(2k+1)}\coloneqq S\circ_rS\circ\cdots\circ_rS$ in Corollary~\ref{cor:comm}, in the case where $T=T_i$ and $S=S_i$ for all $i$.

\begin{Lemma}\label{lam:mixed_suff}
 Let $T^{(2k+1)}$ and $S^{(2k+1)}$ be as above. Suppose that $T$ and $S$ satisfy \eqref{eq:1}--\eqref{eq:3}, and when $k>1$, in addition, satisfy the following relations:
 \begin{align}
& T_{12}S_{13}= S_{13}T_{12},\label{eq:4}\\
& T_{23}S_{34}T_{13}S_{12}= S_{34}T_{24}S_{12}T_{23},\label{eq:5}\\
& T_{12}S_{13}T_{34}S_{23}= S_{23}T_{12}S_{24}T_{34}.\label{eq:6}
 \end{align}
 Graphically, they are
 \begin{equation*}
 \adjustimage{valign=m}{eq4-1.pdf}\ = \ \adjustimage{valign=m}{eq4-2.pdf}\,,\qquad\adjustimage{valign=m}{eq5-1.pdf}\ = \ \adjustimage{valign=m}{eq5-2.pdf}\,, \qquad\adjustimage{valign=m}{eq6-1.pdf}\ = \ \adjustimage{valign=m}{eq6-2.pdf}\,,
\end{equation*}
 respectively.
 Then $T^{(2k+1)}$ and $S^{(2k+1)}$ satisfy the mixed relation.
\end{Lemma}

\begin{proof}
 We treat the $k=3$ case, and the general case follows in the same way.

 The graphical representations of $T^{(7)}$ and $S^{(7)}$ are
 \begin{equation*}
 T^{(7)} = \ \adjustimage{valign=m}{7-gon_t.pdf}\qquad\text{and}\qquad S^{(7)} = \ \adjustimage{valign=m}{7-gon_s.pdf} \end{equation*}
 Now, we need to show the following equality, obtained by substituting $T^{(7)}$ and $S^{(7)}$ into the mixed relation:
 \begin{equation*}
 \adjustimage{valign=m}{7-gon_mixed_lhs1.pdf} \ = \ \adjustimage{valign=m}{7-gon_mixed_rhs1.pdf} \end{equation*}
 Then the proof proceeds graphically as follows, where we apply the relation indicated above each $=$ sign to the highlighted edges:
 \begin{align*}
 &\text{(LHS)} \overset{\substack{\eqref{eq:1}\\\eqref{eq:2}}}{=} \ \adjustimage{valign=m, scale=0.94}{7-gon_mixed_lhs2.pdf}\ &\overset{\substack{\eqref{eq:1}\\\eqref{eq:4}\\\eqref{eq:5}}}{=}& \ \adjustimage{valign=m, scale=0.94}{7-gon_mixed_lhs3.pdf} &\overset{\substack{\eqref{eq:1}\\\eqref{eq:2}}}{=}& \ \adjustimage{valign=m, scale=0.94}{7-gon_mixed_lhs4.pdf}\\
 &\overset{\substack{\eqref{eq:1}\\\eqref{eq:4}\\\eqref{eq:5}}}{=} \ \adjustimage{valign=m, scale=0.94}{7-gon_mixed_lhs5.pdf}\ &\overset{\substack{\eqref{eq:1}\\\eqref{eq:4}\\\eqref{eq:5}}}{=}& \ \adjustimage{valign=m, scale=0.94}{7-gon_mixed_lhs6.pdf}\ &\overset{\substack{\eqref{eq:1}\\\eqref{eq:2}}}{=}& \quad\adjustimage{valign=m, scale=0.94}{7-gon_mixed_lhs7.pdf} \\[10pt]
 &\text{(RHS)} \overset{\substack{\eqref{eq:1}\\\eqref{eq:3}}}{=} \ \adjustimage{valign=m, scale=0.94}{7-gon_mixed_rhs2.pdf}\quad &\overset{\substack{\eqref{eq:1}\\\eqref{eq:4}\\\eqref{eq:6}}}{=}& \ \adjustimage{valign=m, scale=0.94}{7-gon_mixed_rhs3.pdf}\ &\overset{\substack{\eqref{eq:1}\\\eqref{eq:3}}}{=}& \ \adjustimage{valign=m, scale=0.94}{7-gon_mixed_rhs4.pdf}\\
 &\overset{\substack{\eqref{eq:1}\\\eqref{eq:4}\\\eqref{eq:6}}}{=} \ \adjustimage{valign=m, scale=0.94}{7-gon_mixed_rhs5.pdf}\ &\overset{\substack{\eqref{eq:1}\\\eqref{eq:4}\\\eqref{eq:6}}}{=}& \ \adjustimage{valign=m, scale=0.94}{7-gon_mixed_rhs6.pdf}\ &\overset{\substack{\eqref{eq:1}\\\eqref{eq:3}}}{=}& \ \adjustimage{valign=m, scale=0.94}{7-gon_mixed_rhs7.pdf}\!\!\!\!\!\tag*{\qed}
 \end{align*}\renewcommand{\qed}{}
 \end{proof}

\begin{Example}
 Let $H$ be a commutative and cocommutative Hopf algebra with antipode $s$, and $T^{(2k+1)}$ be the solution of $(2k+1)$-gon equation in Example~\ref{ex:ex from comm Hopf}.
 Set $M_S \coloneqq M \circ (s\otimes \operatorname{id}_H)$ and
 \begin{align*}
S^{(2k+1)} \coloneqq \underbrace{\Delta \circ_r M_S \circ_r \Delta \circ_r M_S \circ_r \cdots \circ_{r} M_S}_{2k-2},\quad k>1, \qquad \text{and}\qquad S^{(3)} \coloneqq {\rm id}_H.
 \end{align*}
 Then $S^{(2k+1)}$ is a solution of the dual $(2k+1)$-gon equation. One can also check that~$T^{(3)}$ and~$S^{(3)}$ satisfy \eqref{eq:1}--\eqref{eq:6} in Lemma~\ref{lam:mixed_suff}, and therefore the pair $\bigl(T^{(2k+1)},S^{(2k+1)}\bigr)$ satisfies the mixed relation.
\end{Example}

\appendix
\section{Diagrammatic construction of polygon and simplex equations}\label{sec:pachnerpolygon}
In this appendix, we present a systematic method for constructing (possibly non-constant) polygon and simplex equations in relation to simplices in PL topology.

\subsection{Pachner moves and polygon equations}
This section reviews the relationship between Pachner moves and polygon equations. (See also \cite{kashaev2015realizations,korepanov2024odd,wan2024matrix}.)

\begin{figure}[!ht]
 \centering
 \begin{subcaptionblock}{.4\textwidth}
 \centering
 \includegraphics[width=.9\linewidth]{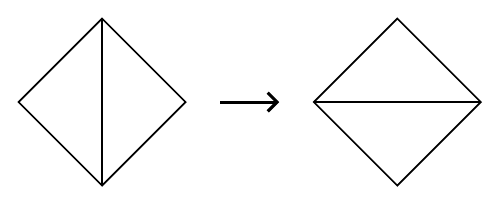}
 \caption*{Pachner $(2,2)$-move}
 \end{subcaptionblock}
 \begin{subcaptionblock}{.4\textwidth}
 \centering
 \includegraphics[width=.8\linewidth]{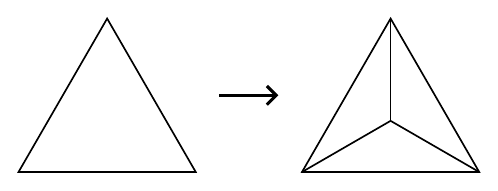}
 \caption*{Pachner $(1,3)$-move}
 \end{subcaptionblock}
 \begin{subcaptionblock}{.4\textwidth}
 \centering
 \includegraphics[width=.9\linewidth]{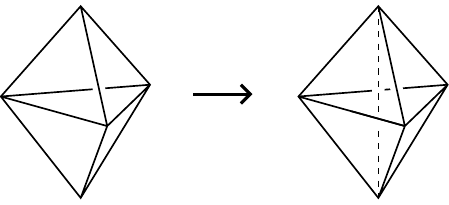}
 \caption*{Pachner $(2,3)$-move}
 \end{subcaptionblock}
 \begin{subcaptionblock}{.4\textwidth}
 \centering
 \includegraphics[width=.9\linewidth]{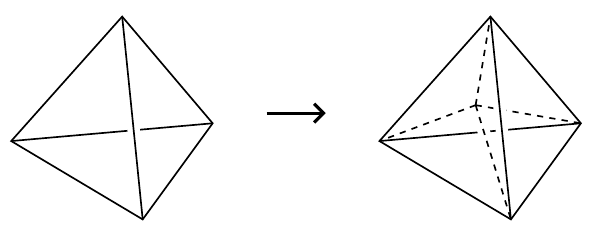}
 \caption*{Pachner $(1,4)$-move}
 \end{subcaptionblock}
 \caption{Pachner moves in two dimensions (top) and three dimensions (bottom).}
 \label{fig:pachnermoves}
\end{figure}

 Let $\Delta^{n+1}=[0,1,\dots, n+1]$ be the standard $(n+1)$-simplex. We write the boundary $\partial\Delta^{n+1}=A\cup B$ of the $(n+1)$-simplex $\Delta^{n+1}$ as the union of two $n$-dimensional subcomplexes $A$ and $B$. Each of $A$ and $B$ is homeomorphic to an $n$-disk, and they share a common $(n-1)$-sphere boundary. Let $i$ and $j$ denote the numbers of $n$-simplices of $A$ and $B$, respectively. A Pachner $(i,j)$-move on an $n$-dimensional manifold is the operation that replaces a subcomplex isomorphic to $A$ with one isomorphic to $B$, glued along the common boundary. Pachner's theorem~\cite{pachner1991pl} states that any two triangulations of a closed $n$-dimensional piecewise-linear manifold are related by a finite sequence of Pachner $(k,n+2-k)$-moves, where $1\le k\le n+1$. Examples are shown in Figure~\ref{fig:pachnermoves}.

\begin{figure}[!t]
 \centering
 \begin{subcaptionblock}{.4\textwidth}
 \centering
 \includegraphics[width=.9\linewidth]{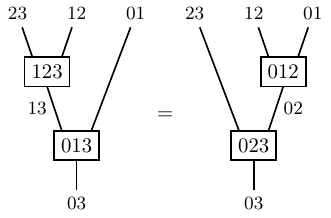}
 \caption*{4-gon equation}
 \end{subcaptionblock}
 \begin{subcaptionblock}{.4\textwidth}
 \centering
 \includegraphics[width=.9\linewidth]{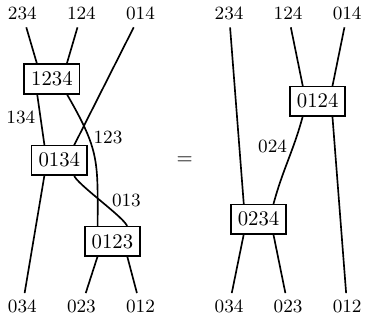}
 \caption*{5-gon (pentagon) equation}
 \end{subcaptionblock}
 \caption{Realizations of 4- and 5-gon equations, read from bottom to top. Here we omit the notations $T$ for maps and $V$ for spaces. Reading from top to bottom yields the dual 4- and 5-gon equations.}
 \label{fig:graphical45goneq}
\end{figure}

Let $V$ be a vector space. We specify the $(n+1)$-simplex by its $n+2$ vertices (equipped with the standard ordering): $\Delta^{n+1}=[0,1,\dots, n+1]$. To each $(n-1)$-simplex $s$ of $\Delta^{n+1}$, we associate a vector space $V(s)=V$, and to each $n$-simplex $p$ of $\Delta^{n+1}$, we associate a linear map
\begin{equation*}
 T(p)\colon \ V(\partial_1 p)\otimes V(\partial_3 p)\otimes\cdots\otimes V\bigl(\partial_{2\lfloor\frac{n+1}{2}\rfloor-1} p\bigr)\to V(\partial_0 p)\otimes V(\partial_2 p)\otimes\cdots\otimes V\bigl(\partial_{2\lfloor\frac{n}{2}\rfloor} p\bigr),
\end{equation*}
 where $\partial_ip \coloneqq [v_0,\dots,\hat{v}_i,\dots,v_n]$ for $p=[v_0\dots,v_n]$ and $0\leq i\leq n$. We draw the map $T(p)$ as below and read it from bottom to top:
\begin{equation*}
 \adjustimage{valign=m}{linearmap.pdf}
\end{equation*}
We now split the vertex set $\{0,1,\dots,n+1\}$ of $\Delta^{n+1}$ into two subsets
\begin{equation*}
 I\coloneqq\left\{0,2,4,\dots,2\left\lfloor\frac{n+1}{2}\right\rfloor\right\}\qquad\text{and}\qquad J\coloneqq\left\{1,3,5,\dots,2\left\lfloor\frac{n}{2}\right\rfloor+1\right\}.
\end{equation*}
We then compose the maps $T(\partial_i\Delta^{n+1})$ for $i\in I$ along the matching labeled outputs and inputs, and arrange the remaining free outputs and free inputs, each in reverse lexicographic order. We do the same for $T(\partial_j\Delta^{n+1})$ for $j\in J$. The resulting equality between these two compositions is called the Pachner $(\lfloor(n+1)/2\rfloor+1,\lfloor n/2\rfloor+1)$-relation, which coincides with the $(n+2)$-gon equation. We think of it as an algebraic realization of the Pachner $(\lfloor(n+1)/2\rfloor+1,\lfloor n/2\rfloor+1)$-move
\begin{equation*}
 \bigcup_{i\in I}\partial_i\Delta^{n+1}\longleftrightarrow\bigcup_{j\in J}\partial_j\Delta^{n+1}.
\end{equation*}
Pictures of Pachner relations in 2- and 3-dimensions, corresponding to 4- and 5-gon equations, respectively, are shown in Figure~\ref{fig:graphical45goneq}.

If we instead associate
\begin{equation*}
 S(p)\colon \ V(\partial_0 p)\otimes V(\partial_1 p)\otimes\cdots\otimes V\bigl(\partial_{2\lfloor\frac{n}{2}\rfloor} p\bigr)\to V(\partial_1 p)\otimes V(\partial_3 p)\otimes\cdots\otimes V\bigl(\partial_{2\lfloor\frac{n+1}{2}\rfloor-1} p\bigr)
\end{equation*}
to each $n$-simplex $p\subseteq\Delta^{n+1}$, the same procedure yields the dual $(n+2)$-gon equation (Figure~\ref{fig:graphical45goneq}).

\begin{Remark}
 Note that, in this case, the maps $T(p)$ (and $S(p)$) are not necessarily the same for all $n$-simplices $p$ of $\Delta^{n+1}$. So such a family is called a non-constant solution of the (dual) $(n+2)$-gon equation. If all the maps $T(p)$ (and $S(p)$) are the same, the solution is constant.
\end{Remark}

\subsection{Simplex equations}

A similar systematic construction of simplex equations is also possible.

\begin{figure}
 \centering
 \begin{subcaptionblock}{.4\textwidth}
 \centering
 \includegraphics[width=.9\linewidth]{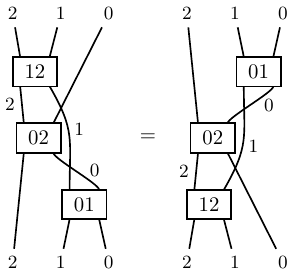}
 \caption*{2-simplex (Yang--Baxter) equation}
 \end{subcaptionblock}
 \begin{subcaptionblock}{.4\textwidth}
 \centering
 \includegraphics[width=.9\linewidth]{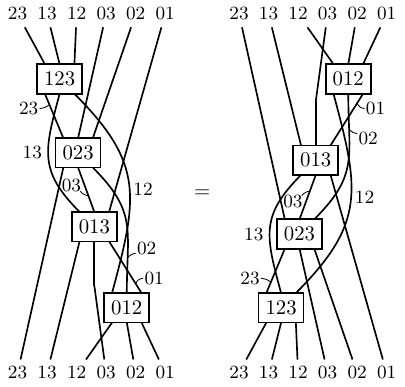}
 \caption*{3-simplex (tetrahedron) equation}
 \end{subcaptionblock}
 \caption{2- and 3-simplex equations.}
 \label{fig:graphical23simpeq}
\end{figure}

Let $\Delta^n=[0,1,\dots,n]$ denote the standard $n$-simplex, and let $V$ be a vector space.
Consider the $n(n+1)/2$-fold tensor product $V^{\otimes n(n+1)/2}$ and label each factor by the $(n-2)$-simplices $\partial_{i,j}\Delta^n\coloneqq [0,\dots,\hat{i},\dots,\hat{j},\dots, n]$. Thus, $\bigotimes_{0\leq i<j\leq n}V(\partial_{i,j}\Delta^n) = V(\partial_{0,1}\Delta^n)\otimes V(\partial_{0,2}\Delta^n)\otimes\cdots\otimes V(\partial_{n-1,n}\Delta^n)$, where the factors are aligned in reverse lexicographic order. To each $(n-1)$-simplex $\partial_i\Delta^{n}$, we associate a linear map
\begin{equation*}
 R(\partial_i\Delta^{n}) \colon \ \bigotimes_{0\leq j \leq n, j\neq i} V(\partial_{j,i} \Delta^n) \to \bigotimes_{0\leq j \leq n, j\neq i} V(\partial_{j,i} \Delta^n).
\end{equation*}
Now, for each $0 \leq i \leq n-1$, we connect the inputs of $R(\partial_i\Delta^{n})$ with outputs of $R(\partial_j\Delta^{n})$ ($i<j$) along the matching labels of $(n-2)$-simplices, yielding the left-hand side of the $n$-simplex equation $\bigotimes_{0\leq i<j\leq n}V(\partial_{i,j}\Delta^n) \to \bigotimes_{0\leq i<j\leq n}V(\partial_{i,j}\Delta^n)$. The right-hand side is obtained by reversing the order of composition, i.e., we connect the outputs of $R(\partial_i\Delta^{n})$ with inputs of $R(\partial_j\Delta^{n})$. See Figure~\ref{fig:graphical23simpeq}.

\begin{Remark}
 As in the polygon equation case, an $n$-simplex equation is constant when the maps $R(p)$ coincide for all $(n-1)$-simplices $p$, and non-constant otherwise.
\end{Remark}

\section[Mixed relation and non-constant versions of Theorems~\ref{thm:fromnton-1} and \ref{thm:fromnton-2}]{Mixed relation and non-constant versions\\ of Theorems~\ref{thm:fromnton-1} and \ref{thm:fromnton-2}}\label{sec:nonconstantthm}

This approach also provides a systematic construction of the mixed relation, from which we obtain non-constant versions of Theorems~\ref{thm:fromnton-1} and \ref{thm:fromnton-2}.

\subsection{Mixed relation}

\begin{figure}[!ht]
 \centering
 \begin{subcaptionblock}{.4\textwidth}
 \centering
 \includegraphics[scale=1]{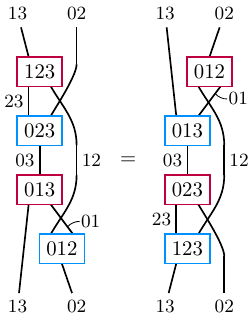}
 \end{subcaptionblock}
 \begin{subcaptionblock}{.5\textwidth}
 \centering
 \includegraphics[scale=1]{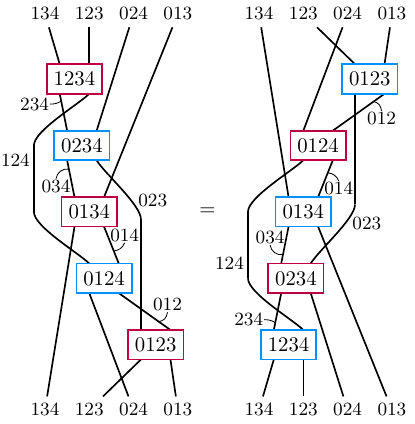}
 \end{subcaptionblock}
 \caption{Mixed relations for 4-gon (left) and 5-gon (right). The blue boxes represent a map~$T$ and the red boxes represent a map $S$ (dual). The free outputs/inputs are labeled with inner simplices, and the inner edges are labeled with the boundary simplices in the Pachner move (compare with Figure~\ref{fig:graphical45goneq}).}
 \label{fig:graphical45gonmixed}
\end{figure}

Let $\bigl\{T\bigl(\partial_0\Delta^{n+1}\bigr),\dots, T\bigl(\partial_{n+1}\Delta^{n+1}\bigr)\bigr\}$ be a (non-constant) solution of the $(n+2)$-gon equation and $\bigl\{S\bigl(\partial_0\Delta^{n+1}\bigr),\allowbreak\dots, S\bigl(\partial_{n+1}\Delta^{n+1}\bigr)\bigr\}$ be a (non-constant) solution of its dual. They are labeled with the $n$-simplices of $\Delta^{n+1}$ as before. Then one side of the mixed relation is obtained by placing, from top to bottom, the maps
\begin{equation*}
 S\bigl(\partial_0\Delta^{n+1}\bigr),\ T\bigl(\partial_1\Delta^{n+1}\bigr),\ \dots,\ \begin{cases}
 S\bigl(\partial_{n+1}\Delta^{n+1}\bigr)& \text{if $n$ is odd,}\\
 T\bigl(\partial_{n+1}\Delta^{n+1}\bigr)& \text{if $n$ is even,}
 \end{cases}
\end{equation*}
composing them along the matching labeled outputs and inputs, and arranging the remaining free outputs and free inputs, each in reverse lexicographic order. The other side is obtained by placing, from bottom to top, the maps
\begin{equation*}
 T\bigl(\partial_0\Delta^{n+1}\bigr),\ S\bigl(\partial_1\Delta^{n+1}\bigr),\ \dots,\ \begin{cases}
 T\bigl(\partial_{n+1}\Delta^{n+1}\bigr) & \text{if $n$ is odd,}\\
 S\bigl(\partial_{n+1}\Delta^{n+1}\bigr) & \text{if $n$ is even,}
 \end{cases}
\end{equation*}
and repeating the same procedure. Examples are shown in Figure~\ref{fig:graphical45gonmixed}.\

\subsection{Non-constant versions of Theorems~\ref{thm:fromnton-1} and \ref{thm:fromnton-2}}

\begin{figure}[t]
 \centering
 \begin{subcaptionblock}{.45\textwidth}
 \centering
 \includegraphics[scale=1]{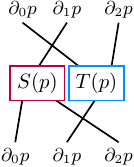}
 \end{subcaptionblock}
 \begin{subcaptionblock}{.45\textwidth}
 \centering
 \includegraphics[scale=1]{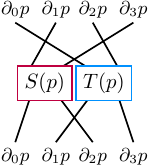}
 \end{subcaptionblock}
 \caption{Left: the solution $R(p)$ of the 3-simplex equation from the (dual) 4-gon equations, where $p$ is any 2-simplex of $\Delta^3$. Right: the solution $R(p)$ of the 4-simplex equation from the (dual) 5-gon equations, where $p$ is any 3-simplex of $\Delta^4$.}
 \label{fig:simp_from_poly1}
\end{figure}

Now, given a mixed pair of (non-constant) solutions $\bigl\{T\bigr(\partial_i\Delta^{n+1}\bigr)\bigr\}_{i=0}^{n+1}$ and $\bigl\{S\bigl(\partial_i\Delta^{n+1}\bigr)\bigr\}_{i=0}^{n+1}$, we construct (non-constant) solutions of the $(n+1)$- and $n$-simplex equations as follows.
For $i=0,1,\dots,n+1$, set
\begin{equation*}
 R\bigl(\partial_i\Delta^{n+1}\bigr) \coloneqq \sigma\circ\bigl(S\bigl(\partial_i\Delta^{n+1}\bigr)\otimes T\bigl(\partial_i\Delta^{n+1}\bigr)\bigr)\circ\tau,
\end{equation*}
where $\sigma$ permutes the labeled outputs of $S\bigl(\partial_i\Delta^{n+1}\bigr)\otimes T\bigl(\partial_i\Delta^{n+1}\bigr)$ to the labeled outputs of $R\bigl(\partial_i\Delta^{n+1}\bigr)$, and $\tau$ permutes the labeled inputs of $R\bigl(\partial_i\Delta^{n+1}\bigr)$ to the labeled inputs of $S\bigl(\partial_i\Delta^{n+1}\bigr)\allowbreak \otimes T\bigl(\partial_i\Delta^{n+1}\bigr)$, i.e.,
\begin{gather*}
 \sigma =
 \begin{pNiceMatrix}[first-row, last-row]
 \rule[-10pt]{0pt}{0pt}\Block{1-4}{\text{outputs of $S$}} &&&& \Block{1-4}{\text{outputs of $T$}}\\
 1 & 3 & \cdots & 2\lfloor\frac{n+1}{2}\rfloor-1 & 0 & 2 & \cdots & 2\lfloor\frac{n}{2}\rfloor \\
 0 & 1 & \multicolumn{5}{c}{\cdots\cdots\cdots\cdots\cdots\cdots\cdots\cdots\cdots} & n\\
 \rule[0pt]{0pt}{15pt}\Block{1-8}{\text{outputs of $R$}} &&&&&&& \\
 \CodeAfter
 \UnderBrace[shorten, yshift=3pt]{1-1}{2-8}{}
 \OverBrace[shorten, yshift=3pt]{1-1}{2-4}{}
 \OverBrace[shorten, yshift=3pt]{1-5}{2-8}{}
 \end{pNiceMatrix},\\
 \tau =
 \begin{pNiceMatrix}[first-row, last-row]
 \rule[-10pt]{0pt}{0pt}\Block{1-8}{\text{inputs of $R$}} &&&&&&&\\
 0 & 1 & \multicolumn{5}{c}{\cdots\cdots\cdots\cdots\cdots\cdots\cdots\cdots} & n\\
 0 & 2 & \cdots & 2\lfloor\frac{n}{2}\rfloor & 1 & 3 & \cdots & 2\lfloor\frac{n+1}{2}\rfloor-1\\
 \rule[0pt]{0pt}{15pt}\Block{1-4}{\text{inputs of $S$}} &&&& \Block{1-4}{\text{inputs of $T$}} \\
 \CodeAfter
 \OverBrace[shorten, yshift=3pt]{1-1}{2-8}{}
 \UnderBrace[shorten, yshift=3pt]{1-1}{2-4}{}
 \UnderBrace[shorten, yshift=3pt]{1-5}{2-8}{}
 \end{pNiceMatrix}.
\end{gather*}
(See Figure~\ref{fig:simp_from_poly1}.) Then $\bigl\{R\bigl(\partial_i\Delta^{n+1}\bigr)\bigr\}_{i=0}^{n+1}$ is a solution of the $(n+1)$-simplex equation.

For the $n$-simplex equation case (Theorem~\ref{thm:fromnton-2}), one has to note that the $n$-gon equation is labeled with the $(n-1)$-simplices of $\Delta^n$, whereas our present solutions $\bigl\{T\bigl(\partial_i\Delta^{n+1}\bigr)\bigr\}_i$ and $\bigl\{S\bigl(\partial_i\Delta^{n+1}\bigr)\bigr\}_i$ of the $(n+1)$-gon equation and its dual, respectively, are labeled with the $n$-simplices of $\Delta^{n+1}$. We construct a solution $\{Q(q)\}$ of the $n$-simplex equation as follows.

\begin{figure}[t]
 \centering
 \includegraphics{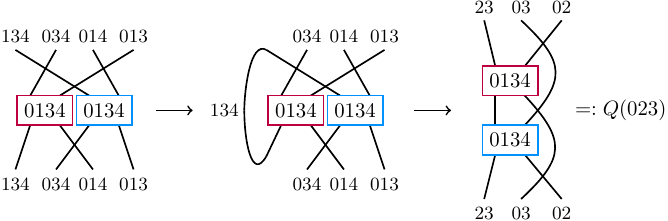}
 \caption{A construction of $Q(q)$ from the 5-gon equation, in which we take the $2$-simplex $q=[023]$ as an example. The first part connects the edges of $R(0134)$ with the label $134$, and the second part arranges the diagram and reindexes the edge labels.}
 \label{fig:simp_from_poly2}
\end{figure}

Let $q=[v_0,v_1,\dots,v_{n-1}]$ be an $(n-1)$-simplex of $\Delta^n$, with which the map $Q(q)$ is labeled. We set $q'\coloneqq[0,v_1+1,v_2+1,\dots,v_{n-1}+1]$ and regard it as an $n$-simplex of $\Delta^{n+1}$. We then connect the input and output of $R(q')$ labeled with $\partial_0q'$ and arrange the diagram so that the order of the rest of the free inputs and outputs is preserved. Finally, we reindex the edge labels via $\partial_1q'\mapsto\partial_0q, \partial_2q'\mapsto\partial_1q,\dots, \partial_nq'\mapsto\partial_{n-1}q$. This yields a map $V^{\otimes n}\to V^{\otimes n}$, which we define as $Q(q)$. The family $\{Q(q)\}_q$, taken over all $(n-1)$-simplices $q$ of $\Delta^n$, now solves the $n$-simplex equation. See Figure~\ref{fig:simp_from_poly2}.

\begin{Remark}
 Note that in both cases, the solutions coincide with those given in Theorems~\ref{thm:fromnton-1} and \ref{thm:fromnton-2} when both $\bigl\{T\bigl(\partial_i\Delta^{n+1}\bigr)\bigr\}_{i=0}^{n+1}$ and $\bigl\{S\bigl(\partial_i\Delta^{n+1}\bigr)\bigr\}_{i=0}^{n+1}$ are constant.
\end{Remark}

\section{Equivalence of two definitions of polygon equations}\label{sec:equivalence}
In this appendix, we explain the correspondence between our definition of polygon equations and the definition given in \cite{dimakis2015simplex}.

In Appendix~\ref{sec:pachnerpolygon}, we discuss the relationship between polygon equations and Pachner moves. Let us return to this discussion and look carefully at how each map $T(p)$ acts on the tensor product of vector spaces $V(s)$.
The domain of the $n$-gon equation, corresponding to the free inputs of the Pachner $(\lfloor(n-1)/2\rfloor+1,\lfloor (n-2)/2\rfloor+1)$-move, is labeled by the $(n-3)$-simplices $\partial_i\partial_j\Delta^{n-1}$, where $i$~is odd, $j$~is even, and $i<j$:
\begin{equation}\label{eq:free_inputs_of_n-gon}
 \bigotimes_{\substack{0\leq i<j\leq n-1\\\text{$i$: odd, $j$: even}}} V\bigl(\partial_i\partial_j\Delta^{n-1}\bigr).
\end{equation}
Then for the construction of the left-hand side of the $n$-gon equation, we first apply the map
\[
T\bigl(\partial_{2\lfloor\frac{n-1}{2}\rfloor}\Delta^{n-1}\bigr),
\]
which replaces the spaces in \eqref{eq:free_inputs_of_n-gon} labeled by
\begin{equation*}
 \partial_1\partial_{2\lfloor\frac{n-1}{2}\rfloor}\Delta^{n-1},\quad \partial_3\partial_{2\lfloor\frac{n-1}{2}\rfloor}\Delta^{n-1},\quad\dots,\quad\partial_{2\lfloor\frac{n-1}{2}\rfloor-1}\partial_{2\lfloor\frac{n-1}{2}\rfloor}\Delta^{n-1}
\end{equation*}
with the spaces labeled by
\begin{equation*}
 \partial_0\partial_{2\lfloor\frac{n-1}{2}\rfloor}\Delta^{n-1},\quad \partial_2\partial_{2\lfloor\frac{n-1}{2}\rfloor}\Delta^{n-1},\quad\dots,\quad \partial_{2\lfloor\frac{n-2}{2}\rfloor}\partial_{2\lfloor\frac{n-1}{2}\rfloor}\Delta^{n-1}.
\end{equation*}
We next apply \smash{$T\bigl(\partial_{2\lfloor\frac{n-1}{2}\rfloor-2}\Delta^{n-1}\bigr)$} and continue in the same way down to $T\bigl(\partial_0\Delta^{n-1}\bigr)$. Finally, we arrive at the codomain, corresponding to the free outputs, which is labeled by the $(n-3)$-simplices $\partial_k\partial_l\Delta^{n-1}$, where $k$ is even, $l$ is odd, and $k<l$:
\begin{equation}\label{eq:free_outputs_of_n-gon}
 \bigotimes_{\substack{0\leq k<l\leq n-1\\\text{$k$: even, $l$: odd}}} V\bigl(\partial_k\partial_l\Delta^{n-1}\bigr).
\end{equation}
Similarly, for the construction of the right-hand side of the $n$-gon equation, we apply
\[
T\bigl(\partial_1\Delta^{n-1}\bigr),\quad T\bigl(\partial_3\Delta^{n-1}\bigr),\quad \dots,\quad T\bigl(\partial_{2\lfloor\frac{n-2}{2}\rfloor+1}\Delta^{n-1}\bigr)
\]
in turn to \eqref{eq:free_inputs_of_n-gon}, and arrive at the same codomain labels~\eqref{eq:free_outputs_of_n-gon}.

Therefore, we obtain the following two chains of transitions of $(n-3)$-simplices, corresponding to the left- and right-hand sides of the $n$-gon equation.
In these chains, the $(n-3)$-simplices are arranged in reverse lexicographic order (with respect to the ordered vertices of $(n-3)$-simplices $[v_0,v_1,\cdots, v_{n-3}]$), and the symbol $\Delta^{n-1}$ is omitted,
\begin{gather*}
 \begin{array}{@{}c@{\,}c@{\,}c@{\,}ccc@{\,}c@{}}
 \begin{array}{@{}c@{}}
 \partial_1\partial_2\\
 \vdots\\
 \partial_1\partial_{2\lfloor\frac{n-1}{2}\rfloor}\\
 \vdots\\
 \partial_{2\lfloor\frac{n-1}{2}\rfloor-1}\partial_{2\lfloor\frac{n-1}{2}\rfloor}\\
 \end{array}
 &
 \xrightarrow{\partial_{2\lfloor\frac{n-1}{2}\rfloor}}
 &
 \begin{array}{c}
 \partial_0\partial_{2\lfloor\frac{n-1}{2}\rfloor}\\
 \partial_1\partial_2\\
 \vdots\\
 \partial_2\partial_{2\lfloor\frac{n-1}{2}\rfloor}\\
 \vdots\\
 \partial_{2\lfloor\frac{n-2}{2}\rfloor}\partial_{2\lfloor\frac{n-1}{2}\rfloor}
 \end{array}
 &
 \xrightarrow{\partial_{2\lfloor\frac{n-3}{2}\rfloor}}
 &
 \cdots
 &
 \xrightarrow{\quad\partial_0\quad}
 &
 \begin{array}{c@{}}
 \partial_0\partial_1\\
 \partial_0\partial_3\\
 \vdots\\
 \partial_{2\lfloor\frac{n-2}{2}\rfloor}\partial_{2\lfloor\frac{n-2}{2}\rfloor+1}
 \end{array}
 \end{array}\\[10pt]
 \begin{array}{@{}c@{\,}c@{\,}c@{\,}ccc@{\,}c@{}}
 \begin{array}{@{}c@{}}
 \partial_1\partial_2\\
 \vdots\\
 \partial_1\partial_{2\lfloor\frac{n-1}{2}\rfloor}\\
 \vdots\\
 \partial_{2\lfloor\frac{n-1}{2}\rfloor-1}\partial_{2\lfloor\frac{n-1}{2}\rfloor}\\
 \end{array}
 &
 \xrightarrow{\quad\partial_1\quad}
 &
 \begin{array}{c}
 \partial_0\partial_1\\
 \partial_1\partial_3\\
 \vdots\\
 \partial_1\partial_{2\lfloor\frac{n-2}{2}\rfloor+1}\Delta^{n-1}\\
 \vdots\\
 \partial_{2\lfloor\frac{n-1}{2}\rfloor-1}\partial_{2\lfloor\frac{n-1}{2}\rfloor}\\
 \end{array}
 &
 \xrightarrow{\quad\partial_3\quad}
 &
 \cdots
 &
 \xrightarrow{\partial_{2\lfloor\frac{n-2}{2}\rfloor+1}}
 &
 \begin{array}{c@{}}
 \partial_0\partial_1\\
 \partial_0\partial_3\\
 \vdots\\
 \partial_{2\lfloor\frac{n-2}{2}\rfloor}\partial_{2\lfloor\frac{n-2}{2}\rfloor+1}
 \end{array}
 \end{array}
\end{gather*}

Let us now compare this definition of polygon equations with the definition given in \cite{dimakis2015simplex}. We first note that the numbering in \cite{dimakis2015simplex} starts from $1$, whereas ours starts from $0$. Thus, to convert their notation to ours, one should subtract $1$ from each index. For example, $[i,j,k]$ in their notation corresponds to $[i-1,j-1,k-1]$ in ours.

In both our definition and that in \cite{dimakis2015simplex}, the $n$-gon equation maps the spaces labeled by the $(n-3)$-simplices $\partial_i\partial_j\Delta^{n-1}$, where $i$ is odd, $j$ is even, and $i<j$, to the spaces labeled by $\partial_k\partial_l\Delta^{n-1}$, where $k$ is even, $l$ is odd, and $k<l$. Moreover, each linear map appearing in the $n$\nobreakdash-gon equation is labeled by an $(n-2)$-simplex, and the labels of its input and output spaces agree in the two definitions.
The only difference lies in how the $(n-3)$-simplices are arranged.
In the definition in \cite{dimakis2015simplex}, the domain of the $n$-gon equation is first arranged in lexicographic order. Each time a linear map acts, it changes the corresponding lexicographically ordered inputs into reverse lexicographically ordered outputs, and the process finally reaches its codomain, which is arranged in reverse lexicographic order. On the other hand, in our definition, the $(n-3)$-simplices are always arranged in reverse lexicographic order. Therefore, by applying the appropriate inversion to each map, we can translate one definition into the other. 
\begin{Proposition}\quad\samepage
 \begin{enumerate}\itemsep=0pt
 \item[$(1)$] A map $T$ satisfies the $($dual$)$ $(2k+1)$-gon equation in {\rm \cite{dimakis2015simplex}} if and only if $T\circ\bar{\sigma}_k$ satisfies the $($dual$)$ $(2k+1)$-gon equation in this paper, where $\bar{\sigma}_k$ is the inversion defined in~\eqref{eq:inversion}.
 \item[$(2)$] A map $T$ satisfies the $2k$-gon $($resp. dual $2k$-gon$)$ equation in {\rm \cite{dimakis2015simplex}} if and only if $T\circ\bar{\sigma}_k$ $($resp. $T\circ\bar{\sigma}_{k-1})$ satisfies the dual $2k$-gon $($resp. $2k$-gon$)$ equation in this paper.
 \end{enumerate}
\end{Proposition}

To obtain the graphical representation of the polygon equation in Section~\ref{sec:polygon_equations} from this Pachner-based formulation, we place each space $V\bigl(\partial_i\partial_j\Delta^{n-1}\bigr)$ ($i$~is odd, $j$~is even, and $i<j$) in the domain at the point $(j/2, (i+1)/2)\in\mathbb{R}^2$. Then on each side, we apply $T\bigl(\partial_i\Delta^{n-1}\bigr)$ successively, building up the graphical representation. Finally, each space $V\bigl(\partial_k\partial_l\Delta^{n-1}\bigr)$ ($k$~is even, $l$~is odd, and $k<l$) in the codomain corresponds to the point $((l+1)/2,(k+2)/2)$.

\begin{Example}
 Here we consider the dual 4-gon and 5-gon cases. These may be compared with \cite[Section~4.2]{dimakis2015simplex}. See also Figure~\ref{fig:graphical45goneq}.

 For the dual 4-gon equation, which corresponds to the 4-gon equation in \cite{dimakis2015simplex}, the chains corresponding to the left- and right-hand sides are:
 \begin{equation*}
 \begin{array}{@{}c@{\;}c@{\;}c@{\;}c@{\;}c}
 \begin{array}{c}
 23\\
 12\\
 01
 \end{array}
 &
 \xrightarrow{123}
 &
 \begin{array}{c}
 13\\
 01
 \end{array}
 &
 \xrightarrow{013}
 &
 \begin{array}{c@{}}
 03
 \end{array}
 \end{array}
 \qquad\text{and}\qquad
 \begin{array}{@{}c@{\;}c@{\;}c@{\;}c@{\;}c}
 \begin{array}{c}
 23\\
 12\\
 01
 \end{array}
 &
 \xrightarrow{012}
 &
 \begin{array}{c}
 23\\
 02
 \end{array}
 &
 \xrightarrow{023}
 &
 \begin{array}{c@{}}
 03
 \end{array},
 \end{array}
 \end{equation*}
 respectively.

 For the 5-gon equation, the chains corresponding to the left- and right-hand sides are:
 \begin{equation*}
 \begin{array}{@{}c@{\;}c@{\;}c@{\;}c@{\;}c@{\;}c@{\;}c}
 \begin{array}{c}
 034\\
 023\\
 012
 \end{array}
 &
 \xrightarrow{0123}
 &
 \begin{array}{c}
 123\\
 034\\
 013
 \end{array}
 &
 \xrightarrow{0134}
 &
 \begin{array}{c}
 134\\
 123\\
 014
 \end{array}
 &
 \xrightarrow{1234}
 &
 \begin{array}{c@{}}
 234\\
 124\\
 014
 \end{array}
 \end{array}
 \qquad\text{and}\qquad
 \begin{array}{@{}c@{\;}c@{\;}c@{\;}c@{\;}c}
 \begin{array}{c}
 034\\
 023\\
 012
 \end{array}
 &
 \xrightarrow{0234}
 &
 \begin{array}{c}
 234\\
 024\\
 012
 \end{array}
 &
 \xrightarrow{0124}
 &
 \begin{array}{c@{}}
 234\\
 124\\
 014
 \end{array},
 \end{array}
 \end{equation*}
 respectively.
\end{Example}

\subsection*{Acknowledgements}

We are grateful to Yuji Terashima for helpful suggestions. We would also like to thank the anonymous referees for their helpful comments, which greatly improved the paper.

\pdfbookmark[1]{References}{ref}
\LastPageEnding

\end{document}